\pdfoutput=1

\documentclass[11pt,twoside,a4paper,cmspaper,final,collab]{cms-tdr}
\def\svnVersion{338444}\def\cmsCernNoTag{CERN-PH-EP/2015-297}\def\cmsCernDate{\today}\def\cmsMessage{Published in Physical Review D as \href{http://dx.doi.org/10.1103/PhysRevD.93.052011}{\doi{10.1103/PhysRevD.93.052011.}}}

\begin{document}\cmsNoteHeader{EXO-12-054}

\hyphenation{had-ron-i-za-tion}
\hyphenation{cal-or-i-me-ter}
\hyphenation{de-vices}
\RCS$Revision: 331089 $
\RCS$HeadURL: svn+ssh://svn.cern.ch/reps/tdr2/papers/EXO-12-054/trunk/EXO-12-054.tex $
\RCS$Id: EXO-12-054.tex 331089 2016-03-07 16:34:04Z rewang $
\newcommand{\ZZ}{\ensuremath{\cPZ\cPZ}\xspace}
\newcommand{\WZ}{\ensuremath{\PW\cPZ}\xspace}
\newcommand{\WW}{\ensuremath{\PW\PW}\xspace}
\newcommand{\Zjets}{\ensuremath{\cPZ\text{+jets}}\xspace}
\newcommand{\Wjets}{\ensuremath{\PW\text{+jets}}\xspace}
\renewcommand{\Z}{\ensuremath{\cPZ}\xspace}
\newcommand{\W}{\ensuremath{\PW}\xspace}
\newcommand{\tw}{\ensuremath{\cPqt\PW}\xspace}
\newcommand{\dyll}{\ensuremath{\Z/\gamma^*\to\ell^+\ell^-}\xspace}
\newcommand{\mt}{\ensuremath{m_{\mathrm{T}}}\xspace}
\newcommand{\mll}{\ensuremath{m_{\ell\ell}}\xspace}

\newcolumntype{.}{D{.}{.}{-1}}
\newlength\cmsFigWidth
\ifthenelse{\boolean{cms@external}}{\setlength\cmsFigWidth{0.85\columnwidth}}{\setlength\cmsFigWidth{0.4\textwidth}}
\ifthenelse{\boolean{cms@external}}{\providecommand{\cmsLeft}{top}}{\providecommand{\cmsLeft}{left}}
\ifthenelse{\boolean{cms@external}}{\providecommand{\cmsRight}{bottom}}{\providecommand{\cmsRight}{right}}
\ifthenelse{\boolean{cms@external}}{\providecommand{\cmsLLeft}{Top}}{\providecommand{\cmsLLeft}{Left}}
\ifthenelse{\boolean{cms@external}}{\providecommand{\cmsRRight}{Bottom}}{\providecommand{\cmsRRight}{Right}}
\ifthenelse{\boolean{cms@external}}{\providecommand{\CL}{C.L.\xspace}}{\providecommand{\CL}{CL\xspace}}
\ifthenelse{\boolean{cms@external}}{\providecommand{\NA}{\ensuremath{\cdots}}\xspace}{\providecommand{\NA}{---\xspace}}
\ifthenelse{\boolean{cms@external}}{\providecommand{\lastfig}{bottom\xspace}}{\providecommand{\lastfig}{right-hand\xspace}}
\cmsNoteHeader{EXO-12-054}
\title{\texorpdfstring{Search for dark matter and unparticles produced in association with a \Z boson in proton-proton collisions at $\sqrt{s} = 8 \TeV$}{Search for dark matter and unparticles produced in association with a Z boson in proton-proton collisions at sqrt(s) = 8 TeV}}

\date{\today}

\abstract{A search for evidence of particle dark matter (DM) and unparticle production at the LHC has been performed using events
containing two charged leptons, consistent with the decay of a \Z boson, and large missing transverse momentum.
This study is based on data collected with the CMS detector corresponding to
an integrated luminosity of 19.7\fbinv of $\Pp\Pp$ collisions at the LHC at a center-of-mass energy of 8\TeV.
No significant excess of events is observed above the number expected from the standard model contributions.
The results are interpreted in terms of 90\% confidence level limits on the DM-nucleon scattering cross section,
as a function of the DM particle mass, for both spin-dependent and spin-independent scenarios.
Limits are set on the effective cutoff scale $\Lambda$, and
on the annihilation rate for DM particles, assuming that their branching fraction to quarks is 100\%.
Additionally, the most stringent 95\% confidence level limits to date on the unparticle model
parameters are obtained.}

\hypersetup{%
pdfauthor={CMS Collaboration},%
pdftitle={Search for dark matter and unparticles produced in association with a Z boson in proton-proton collisions at sqrt(s) = 8 TeV},%
pdfsubject={CMS},%
pdfkeywords={CMS, physics, exotica, dark matter, unparticles}}

\maketitle

\section{Introduction}
\label{sec:intro}
\begin{figure}[htbp]
\centering
\includegraphics[width=0.3\textwidth]{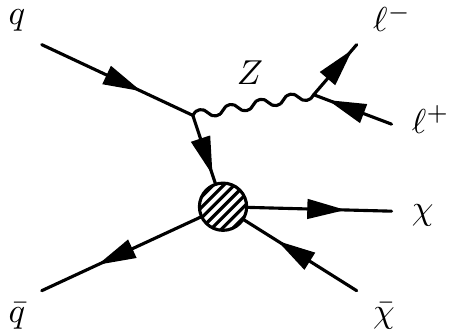}
\includegraphics[width=0.3\textwidth]{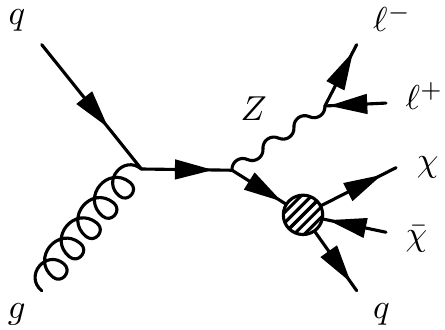}
\includegraphics[width=0.3\textwidth]{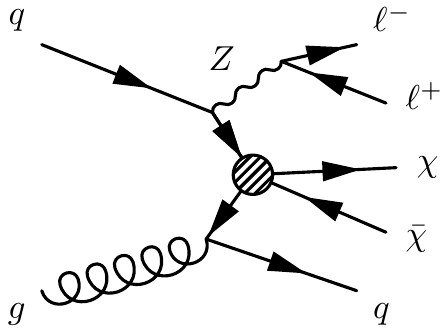}
\caption{The principal Feynman diagrams for the production of DM pairs in association with a \Z  boson.
	In the middle and \lastfig diagrams an additional quark is produced.
	The hatched circles indicate the interaction modeled with an effective field theory.}
\label{fig:DMFeyn}
\end{figure}

Ample evidence from astrophysical measurements supports the existence of dark matter (DM),
which is assumed to be responsible for galactic gravitation that cannot be attributed to baryonic matter~\cite{Tucker:1996sk,LopezCorredoira:1999ff,Clowe:2006eq}.
Recent DM searches have exploited a number of methods including direct detection~\cite{Archambault:2012pm,Felizardo:2011uw,Behnke:2012ys,IceCube:2011aj,
Agnese:2013jaa,Cushman:2013zza,Agnese:2014aze,Akerib:2013tjd},
indirect detection~\cite{Ackermann:2011wa,Buckley:2013bha},
and particle production at colliders~\cite{Khachatryan:2014rra,Aad:2015zva,CMSMonoPhoton8TeV,Aad:2014tda,
Khachatryan:2014tva,Aad:2014vea,ATLASMonoWZhad8TeV,ATLAS:2014wra,Aad:2014wza,Aad:2015yga,Khachatryan:2014uma,Khachatryan:2015nua,ATLASMonoZ8TeV}.
The currently favored possibility is that DM may take the form of weakly interacting massive particles.
The study presented here considers a mechanism for producing such particles at the CERN LHC~\cite{Carpenter:2012rg}.
In this scenario, a \Z boson, produced
in pp collisions, recoils against a pair of DM particles, $\chi\overline\chi$. The \Z boson subsequently decays into
two charged leptons ($\ell^{+}\ell^{-}$, where $\ell=\Pe$ or $\Pgm$) producing a clean dilepton signature together with missing transverse momentum
due to the undetected DM particles.
In this analysis, the DM particle $\chi$ is assumed to be a Dirac fermion or a complex scalar particle of which the
coupling to standard model (SM) quarks $\cPq$ can be described by one of the effective interaction terms~\cite{Goodman:2010ku}:
\begin{align*}
&\text{Vector, spin independent}  \mathrm{(D5): }& \frac{\bar{\chi}\gamma^{\mu}\chi\bar{q}\gamma_{\mu}q}{\Lambda^2}; \\
&\text{Axial vector, spin dependent} \mathrm{(D8): }& \frac{\bar{\chi}\gamma^{\mu}\gamma^5\chi\bar{q}\gamma_{\mu}\gamma^5q}{\Lambda^2}; \\
&\text{Tensor, spin dependent} \mathrm{(D9): } & \frac{\bar{\chi}\sigma^{\mu\nu}\chi\bar{q}\sigma_{\mu\nu}q}{\Lambda^2}; \\
&\text{Vector, spin independent} \mathrm{(C3): } & \frac{\chi^{\dagger}\overset{\leftrightarrow}{\partial_{\mu}}\chi\bar{q}\gamma^{\mu}q}{\Lambda^2};
\end{align*}
where $\Lambda$ parametrizes the effective cutoff scale for interactions between DM particles and quarks.
The operators denoted by D5, D8, and D9 couple to Dirac fermions, while C3 couples to complex scalars.
The corresponding Feynman diagrams for production of a DM pair with a \Z  boson and up to one jet are shown in Fig.~\ref{fig:DMFeyn}.
A search similar to the one presented here has been performed by the ATLAS Collaboration~\cite{ATLASMonoZ8TeV},
where the DM particle is assumed to be a Dirac fermion and couples to either vector bosons or quarks.

\begin{figure}[h!tb]
\centering
   \includegraphics[width=0.35\textwidth]{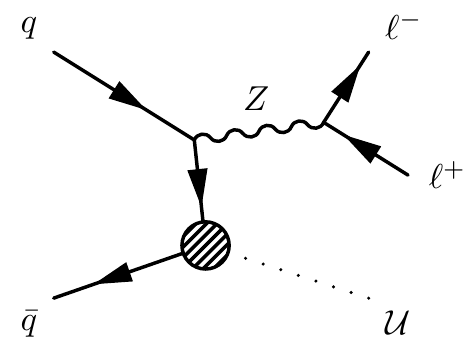}
   \includegraphics[width=0.35\textwidth]{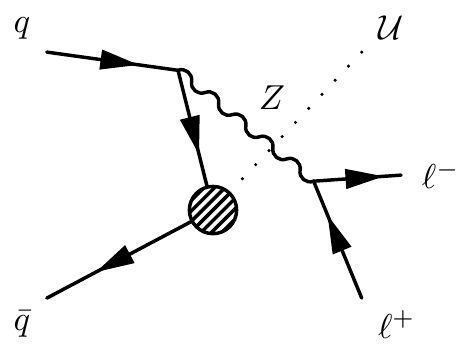}
  \caption{Feynman diagrams for unparticle production in association with a \Z  boson.
	   The hatched circles indicate the interaction modeled with an effective field theory.}
\label{fig:Unpart}
\end{figure}

The unparticle physics concept~\cite{Georgi:2007ek,Georgi:2007si,Cheung:2007zza,Cheung:2007ap} is particularly
interesting because it is based on scale invariance, which is anticipated in many beyond-the-SM physics scenarios~\cite{Kang:2014cia,Rinaldi:2014gha,Cheng:1988zx}.
The unparticle stuff of the scale-invariant sector appears as a noninteger number of invisible massless particles.
In this scenario, the SM is extended by introducing a scale-invariant Banks-Zaks ($\mathcal{BZ}$) field, which has a nontrivial infrared fixed point~\cite{Banks:1981nn}.
This field can interact with SM particles by exchanging heavy particles with a
high mass scale $M_\mathcal{U}$. Below this mass scale, the coupling is nonrenormalizable and the interaction
is suppressed by powers of $M_\mathcal{U}$.
The interaction Lagrangian density can be expressed as
$\mathcal L_{\text{int}} = \mathcal{O}_{\mathrm{SM}}\mathcal{O}_\mathcal{BZ}/M_\mathcal{U}^k,$ where $\mathcal{O}_{\mathrm{SM}}$ is the
operator for the SM field with scaling dimension $d_{\mathrm{SM}}$, $\mathcal{O}_\mathcal{BZ}$ is the operator for
the $\mathcal{BZ}$ field with scaling dimension $d_\mathcal{BZ}$, and $k=d_{\mathrm{SM}}+d_\mathcal{BZ}-4>0$.
At an energy scale of $\Lambda_\mathcal{U}$, dimensional transmutation is induced by renormalization effects in
the scale-invariant $\mathcal{BZ}$ sector, and $\mathcal{O}_\mathcal{BZ}$ can be matched to a new set of operators
below $\Lambda_\mathcal{U}$ with the interaction form
\begin{equation}
\mathcal L_{\text{int}}^{\text{eff}} = C_\mathcal{U}
		\frac{\Lambda_\mathcal{U}^{d_\mathcal{BZ}-d_\mathcal{U}}}
		{M_\mathcal{U}^{k}}\mathcal{O}_{\mathrm{SM}}\mathcal{O}_\mathcal{U}
	= \frac{\lambda}{\Lambda_\mathcal{U}^{d_\mathcal{U}}} \mathcal{O}_{\mathrm{SM}}\mathcal{O}_\mathcal{U},
\label{eq:Lagunparticle}
\end{equation}
in which $C_\mathcal{U}$ is a normalization factor fixed by the matching, $d_\mathcal{U}$
represents the possible noninteger scaling dimension of the unparticle operator $\mathcal{O}_\mathcal{U}$,
and the parameter $\lambda = C_\mathcal{U}\Lambda^{d_\mathcal{BZ}}_\mathcal{U}/M_\mathcal{U}^{k}$
is a measure of the coupling between SM particles and unparticles. In general, an unparticle does
not have a fixed invariant mass but has instead a continuous mass spectrum, and its real
production in low energy processes described by the effective field theory in Eq.~(\ref{eq:Lagunparticle})
can give rise to an excess of missing energy because of the possible nonintegral values of
the scaling dimension $d_\mathcal{U}$.
In the past, the reinterpretation~\cite{Kathrein:2010ej} of LEP single-photon data has been used to set unparticle limits.
A recent search for unparticles at CMS~\cite{Khachatryan:2014rra} in monojet final states has shown no evidence for their existence.
In this paper, a scalar unparticle with real emission is considered, and the
scaling dimension $d_\mathcal{U}>1$ is constrained by the unitarity condition. Figure~\ref{fig:Unpart}
shows the two tree-level diagrams considered in this paper for the production of unparticles associated with a \Z  boson.

Both the DM and unparticle scenarios considered in this analysis produce a dilepton ($\Pep\Pem$ or $\Pgmp\Pgmm$) signature
consistent with a \Z  boson, together with a large magnitude of missing transverse momentum.
The analysis is based on the full data set recorded by the CMS detector in 2012,
which corresponds to an integrated luminosity of $19.7\pm0.5$\fbinv~\cite{CMS-PAS-LUM-13-001}
at a center-of-mass energy of 8\TeV.

\section{CMS detector}
\label{sec:cms}
The CMS detector is a multipurpose apparatus well suited to study high transverse momentum (\pt)
physics processes in pp collisions. The central feature of the CMS apparatus is a superconducting solenoid
of 6\unit{m} internal diameter, providing a magnetic field of 3.8\unit{T}.
Within the superconducting solenoid volume are a silicon pixel and strip tracker,
a lead tungstate crystal electromagnetic calorimeter (ECAL), and a brass and
scintillator hadron calorimeter (HCAL), each composed of a barrel and two end cap sections.
Forward calorimeters extend the pseudorapidity~\cite{CMSdetector} coverage
provided by the barrel and end cap detectors. The electromagnetic calorimeter
consists of 75\,848 lead tungstate crystals, which provide coverage in
pseudorapidity $\abs{ \eta }< 1.479 $ in a barrel region and $1.48 <\abs{ \eta } < 3.00$
in two end cap regions (EE). A preshower detector consisting of two planes
of silicon sensors interleaved with a total of $3 X_0$ of lead is located in front of the EE.
The electron momentum is estimated by combining the energy measurement
in the ECAL with the momentum measurement in the tracker.
The momentum resolution for electrons with $\pt \approx 45$\GeV from $\Z \to \Pe \Pe$
decays ranges from 1.7\% for nonshowering electrons in the barrel region to 4.5\% for
showering electrons in the end caps~\cite{Khachatryan:2015hwa}.
Muons are measured in the pseudorapidity range $\abs{\eta}< 2.4$,
with gas-ionization detectors embedded in the steel flux-return yoke outside the solenoid.
The muon detection planes are made using three technologies: drift tubes, cathode strip chambers, and
resistive plate chambers.
Matching muons to tracks measured in the silicon tracker
results in a relative transverse momentum resolution for muons with $20 <\pt < 100\GeV$ of 1.3--2.0\%
in the barrel and better than 6\% in the end caps. The \pt resolution in the
barrel is better than 10\% for muons with \pt up to 1\TeV~\cite{Chatrchyan:2012xi}.
The first level of the CMS trigger system, composed of custom hardware processors,
uses information from the calorimeters and muon detectors to select the most interesting
events, in a fixed time interval of less than 4\mus. The high-level trigger
processor farm further decreases the event rate from around 100\unit{kHz}
to less than 1\unit{kHz}, before data storage.
A more detailed description of the CMS detector, together with a definition of the coordinate
system used and the relevant kinematic variables, can be found in Ref.~\cite{CMSdetector}.
Variables of particular relevance to the present analysis are the missing transverse momentum
vector \ptvecmiss and the magnitude of this quantity, \ETm. The quantity \ptvecmiss is defined as
the projection on the plane perpendicular to the beams of the negative vector sum of the momenta of
all reconstructed particles in an event.

\section{Simulation}
\label{sec:simulation}
\begin{figure}[htbp]
\centering
\includegraphics[width=0.48\textwidth]{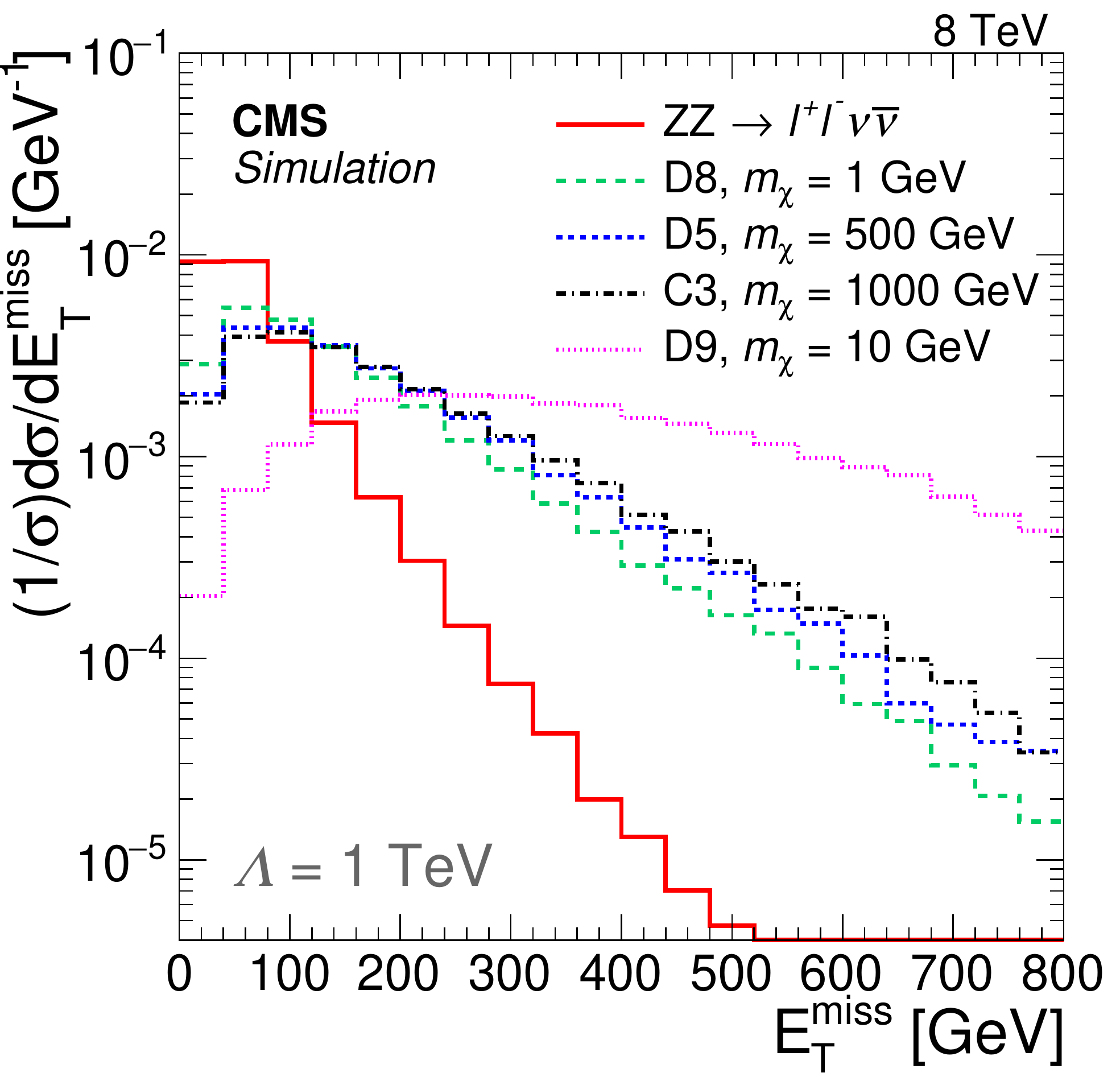}
\includegraphics[width=0.48\textwidth]{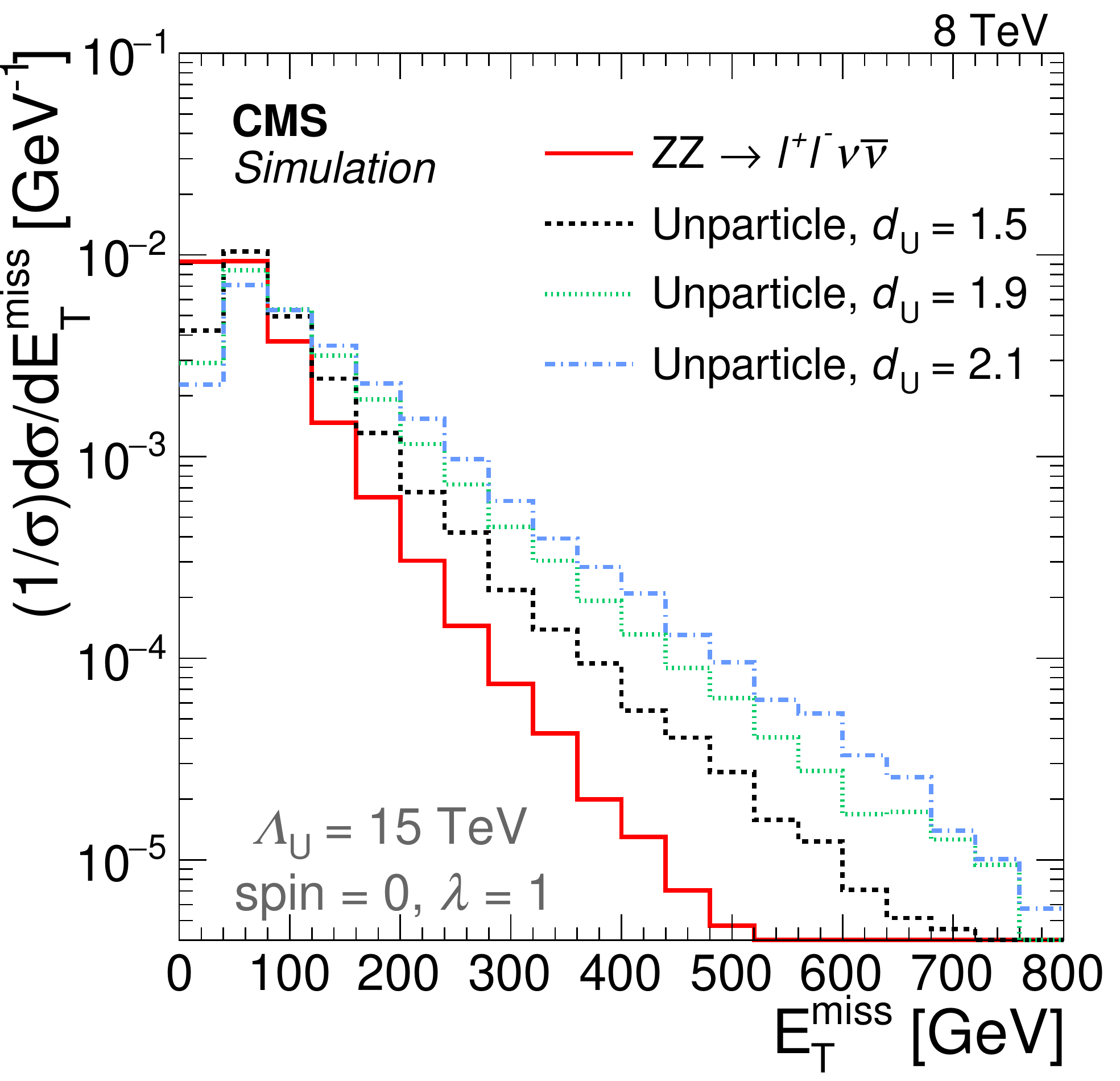}
\caption{The distribution in \ETm at the generator
	level, for DM (\cmsLeft) and unparticle (\cmsRight) scenarios. The DM curves are
	shown for different $m_\chi$ with vector (D5), axial-vector (D8), and
	tensor (D9) coupling for Dirac fermions, and vector (C3) coupling
	for complex scalar particles.
	The unparticle curves have the scalar unparticle coupling $\lambda$ between unparticle and SM
	fields set to 1, with the scaling dimension $d_\mathcal{U}$ ranging from 1.5 to 2.1.
	The SM background $\Z\Z\to\ell^{-}\ell^{+}\PGn\PAGn$ is shown as a red solid curve.
}
\label{fig:genlevel_met}
\end{figure}

Samples of simulated DM particle events are generated using \MADGRAPH5.2.1~\cite{Alwall:2014hca} matched to
\PYTHIA6.4.26~\cite{Sjostrand:2006za} using tune Z2* for parton showering and hadronization.
The \PYTHIA6 Z2* tune uses the CTEQ6L~\cite{Pumplin:2002vw} parton distribution set.
This tune is derived from the  Z1 tune~\cite{Field:2010bc}, which is based on CTEQ5L.
The effective cutoff scale $\Lambda$ is set to 1\TeV. The events for the unparticle
models are generated with \PYTHIA8.1~\cite{Sjostrand:2007gs,Ask:2008fh,Ask:2009pv}
assuming a renormalization scale $\Lambda_\mathcal{U}=15\TeV$,
using tune 4C~\cite{Corke:2010yf} for parton showering and
hadronization.
We evaluate other values of $\Lambda_\mathcal{U}$ by rescaling the cross sections as needed.
The parameter $\Lambda_\mathcal{U}$ acts solely as a scaling factor
for the cross section and does not influence the kinematic distributions of unparticle production~\cite{Ask:2009pv}.
Figure~\ref{fig:genlevel_met} shows the distribution
of \ETm
at the generator level for both DM and unparticle production.
In the unparticle scenario, the events with larger scaling dimension $d_\mathcal{U}$ tend to have a broader \ETm distribution.
For DM production, the shape of the \ETm is similar for couplings D5, D8, and C3, where the vector or axial vector
couplings tend to produce nearly back-to-back DM particles. This configuration is less strongly
favored for the tensor couplings, and thus the D9 couplings show a much broader \ETm distribution.

The \POWHEG2.0~\cite{Nason:2004rx,Frixione:2007vw,Alioli:2010xd,Re:2010bp,Alioli:2011as} event generator is used to produce samples of
events for the $\ttbar$ and $\tw$ background processes.
The $\ZZ$, $\WZ$, and Drell--Yan (DY, $\dyll$)
processes are generated using the \MADGRAPH5.1.3~\cite{Alwall:2007st} event generator.
The default set of parton distribution functions (PDFs) CTEQ6L~\cite{Lai:2010nw} is used
for generators that are leading order (LO) in $\alpha_s$,
while the CT10~\cite{Lai:2010vv} set is used for next-to-leading-order (NLO) generators.
The NLO calculations are used for background cross sections, whereas only LO calculations are available for the signal processes.
For all Monte Carlo (MC) samples, the detector response is simulated using a detailed
description of the CMS detector, based on the \GEANTfour
package~\cite{Agostinelli:2002hh}. Minimum bias events are superimposed on the
simulated events to emulate the additional pp interactions per bunch crossing
(pileup). All MC samples are corrected to reproduce the
pileup distribution as measured in the data. The average number of pileup events
per proton bunch crossing is about 20 for the 2012 data sample.

\section{Event reconstruction}
\label{sec:object}
Events are collected by requiring dilepton ($\Pe\Pe$ or $\Pgm\Pgm$) triggers with thresholds of
$\pt > 17$ and 8\GeV for the leading and subleading leptons, respectively.
Single-lepton triggers with thresholds of $\pt > 27\,(24)\GeV$ for electrons (muons)
are also included to recover residual trigger
inefficiencies. Prior to the selection of leptons, a primary vertex must be
selected as the event vertex. The vertex with largest value of $\sum \pt^2$
for the associated tracks is selected. Simulation studies show that this requirement
correctly selects the event vertex in more than 99\% of
both signal and background events. The lepton candidate tracks are
required to be compatible with the event vertex.

A particle-flow (PF) event algorithm~\cite{CMS-PAS-PFT-09-001,CMS-PAS-PFT-10-001} reconstructs and identifies each individual particle with an optimized combination of information from the various elements of the CMS detector. The energy of photons is directly obtained from the ECAL measurement, corrected for zero-suppression effects. The energy of electrons is determined from a combination of the electron momentum at the event vertex as determined by the tracker, the energy of the corresponding ECAL cluster, and the energy sum of all bremsstrahlung photons spatially compatible with originating from the electron track. The energy of muons is obtained from the curvature of the corresponding track. The energy of charged hadrons is determined from a combination of its momentum measured in the tracker and the matching ECAL and HCAL energy deposits, corrected for zero-suppression effects and for the response function of the calorimeters to hadronic showers. Finally, the energy of neutral hadrons is obtained from the corresponding corrected ECAL and HCAL energy.

Electron candidates are reconstructed using two
algorithms~\cite{Khachatryan:2015hwa}: in the first, energy clusters in
the ECAL are matched to signals in the silicon tracker, and in the second,
tracks in the silicon tracker are matched to ECAL clusters. The electron
candidates used in the analysis are required to be reconstructed by both algorithms.
To reduce the electron misidentification rate, the candidates have to satisfy
additional identification criteria that are based on the shape of the
electromagnetic shower in the ECAL.
In addition, the electron track is required to originate from the event vertex and to match the shower cluster in the ECAL.
Electron candidates with an ECAL cluster in the
transition region between the ECAL barrel and end cap ($1.44 < \abs{\eta} < 1.57$) are
rejected because the reconstruction of an electron object in this region is not optimal.
Candidates that are identified as coming from photon conversions~\cite{Khachatryan:2015hwa} in the
detector material are explicitly removed.

Muon candidate reconstruction is also based on two algorithms: in the first,
tracks in the silicon tracker are matched with at least one muon segment in any detector plane of the muon system,
and in the second algorithm, a combined fit is performed to hits in
both the silicon tracker and the muon system~\cite{Chatrchyan:2012xi}.
The muon candidates in this analysis are required to be reconstructed
by both algorithms and to be further identified as muons by the PF algorithm.
To reduce the muon misidentification rate, additional identification
criteria are applied based on the number of space points measured in the tracker and in
the muon system, the fit quality of the muon track, and its consistency with the event vertex location.

Leptons produced in the decay of \Z bosons are expected to be isolated from hadronic activity in
the event. Therefore, an isolation requirement is applied based on the sum of the momenta of the PF candidates
found in a cone of radius $R=\sqrt{\smash[b]{(\Delta\eta)^2+(\Delta\phi)^2}} = 0.4$ around
each lepton, where $\phi$ is the azimuthal angle.
The isolation sum is required to be smaller than 15\% (20\%) of the $\pt$ of the electron (muon).
To correct for the contribution to the isolation sum from pileup interactions
and the underlying event, a median energy density ($\rho$) is determined on an
event-by-event basis using the method described in Ref.~\cite{FASTJET}.
For each electron, the mean energy deposit in the isolation cone of the electron,
coming from other pp collisions in the same bunch crossing, is estimated
following the method described in Ref.~\cite{Khachatryan:2015hwa}, and subtracted from the isolation sum.
For muon candidates, only charged tracks associated with the event vertex are included.
The sum of the \pt for charged particles not associated with the event vertex
in the cone of interest is rescaled by a factor corresponding to the average neutral
to charge energy densities in jets and subtracted from the isolation sum.

Jets are reconstructed from PF candidates
by using the anti-\kt clustering algorithm~\cite{Cacciari:2008gp}
with a distance parameter of 0.5, as implemented in the {\FASTJET}
package~\cite{Cacciari:2011ma,Cacciari:2006gp}.
Jets are found over the full calorimeter acceptance, $\abs{\eta} < 5$.
The jet momentum is defined as
the vector sum of all particle momenta assigned to the jet and is found in
the simulation to be within 5\% to 10\% of the true hadron-level momentum over the whole \pt
range and detector acceptance. An overall energy subtraction is applied to
correct for the extra energy clustered in jets due to pileup, following the procedure described
in Ref.~\cite{Chatrchyan:2011ds}.
In the subtraction, the charged
particle candidates associated with secondary vertices reconstructed in the event
are also included.  Other jet energy scale corrections applied are derived from
simulation, and are confirmed
by measurements of the energy balance in dijet and $\gamma$+jets events.

\section{Event selection}
\label{sec:selection}
An initial preselection with a large yield is used to validate the background model and is followed by a final selection
that is designed to give maximal sensitivity to the signal.
Selected events are required to have exactly two well-identified, isolated leptons with the same flavor
and opposite charge ($\Pep\Pem$ or $\Pgmp\Pgmm$), each with $\pt > 20$\GeV.
The invariant mass of the lepton pair is required to be within
$\pm$10\GeV of the nominal mass of the $\Z$ boson. Only leptons within the pseudorapidity range of
$\abs{\eta}<2.4\, (2.5)$ for muons (electrons) are considered.
To reduce the background from the $\W\Z$ process where the \W boson decays leptonically, events are removed if an additional
electron or muon is reconstructed with $\pt > 10\GeV$. As a very loose preselection
requirement, the dilepton transverse momentum ($\pt^{\ell\ell}$) is required to be
larger than 50\GeV to reject the bulk of DY background events.

Since only a small amount of hadronic activity is expected in the final state of both DM and
unparticle events, any event having two or more jets with $\pt>30\GeV$ is rejected.
Top quark decays, which always involve the emission of $\cPqb$ quarks, are further suppressed
with the use of techniques based on soft-muon and $\cPqb$-jet tagging. The rejection of events with
soft muons having $\pt >3\GeV$ reduces the background from semileptonic $\cPqb$ decays.
The $\cPqb$-jet tagging technique employed is based
on the ``combined secondary vertex'' algorithm~\cite{Chatrchyan:2012jua,CMS-PAS-BTV-13-001}.
This algorithm selects a group of tracks forming a secondary vertex within a jet and generates
a likelihood discriminant to distinguish between $\cPqb$ jets and jets originating from light quarks, gluons, or charm quarks.
The applied threshold provides, on average, 80\% efficiency for tagging jets originating from $\cPqb$ quarks
and 10\% probability of light-flavor jet misidentification.
The \cPqb-tagged jet is required to have $\pt>20\GeV$ and to be reconstructed within the tracker acceptance volume
($\abs{\eta} < 2.5$).

\begin{figure}[htb!]
\centering
\includegraphics[width=0.48\textwidth]{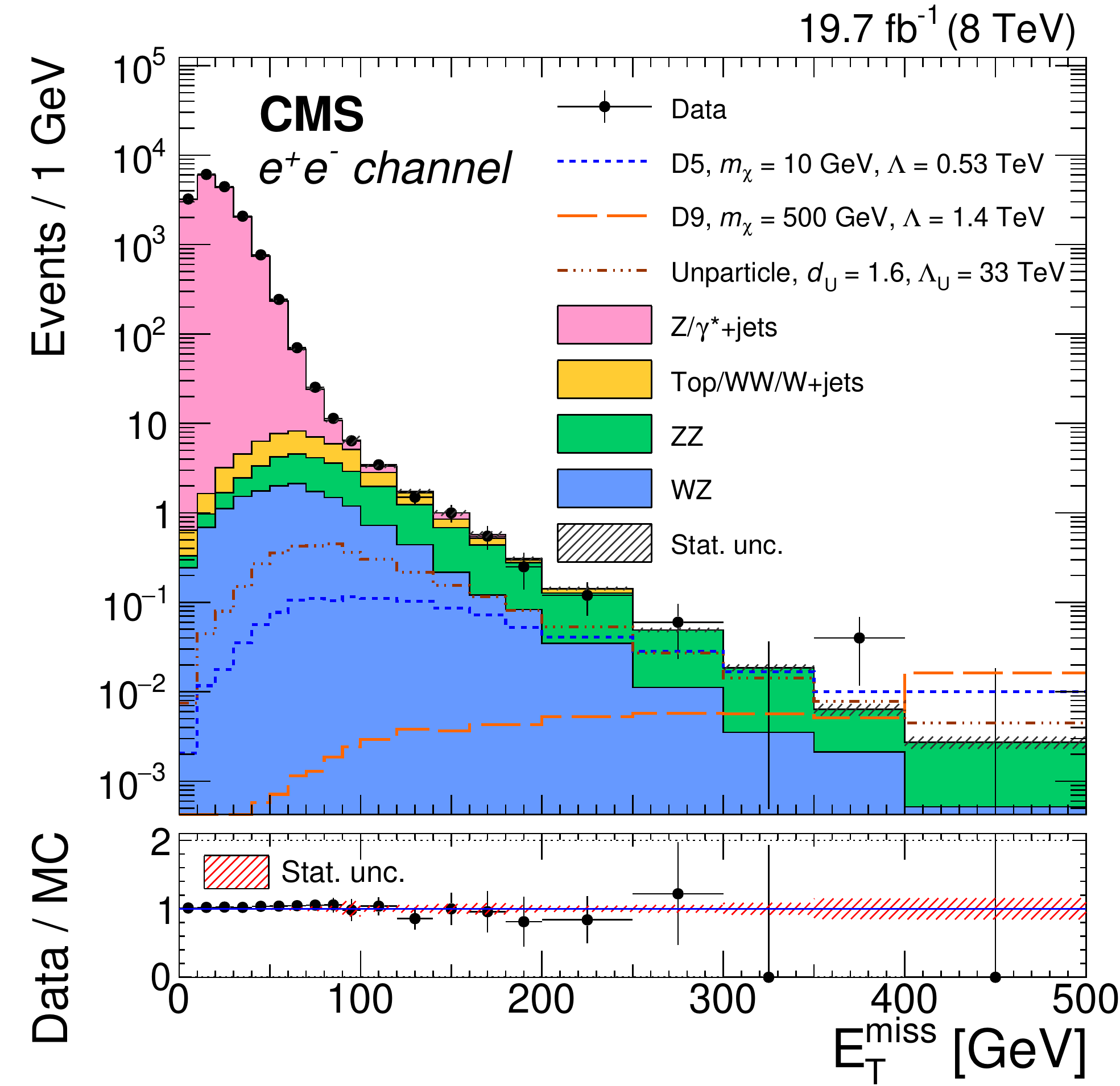}
\includegraphics[width=0.48\textwidth]{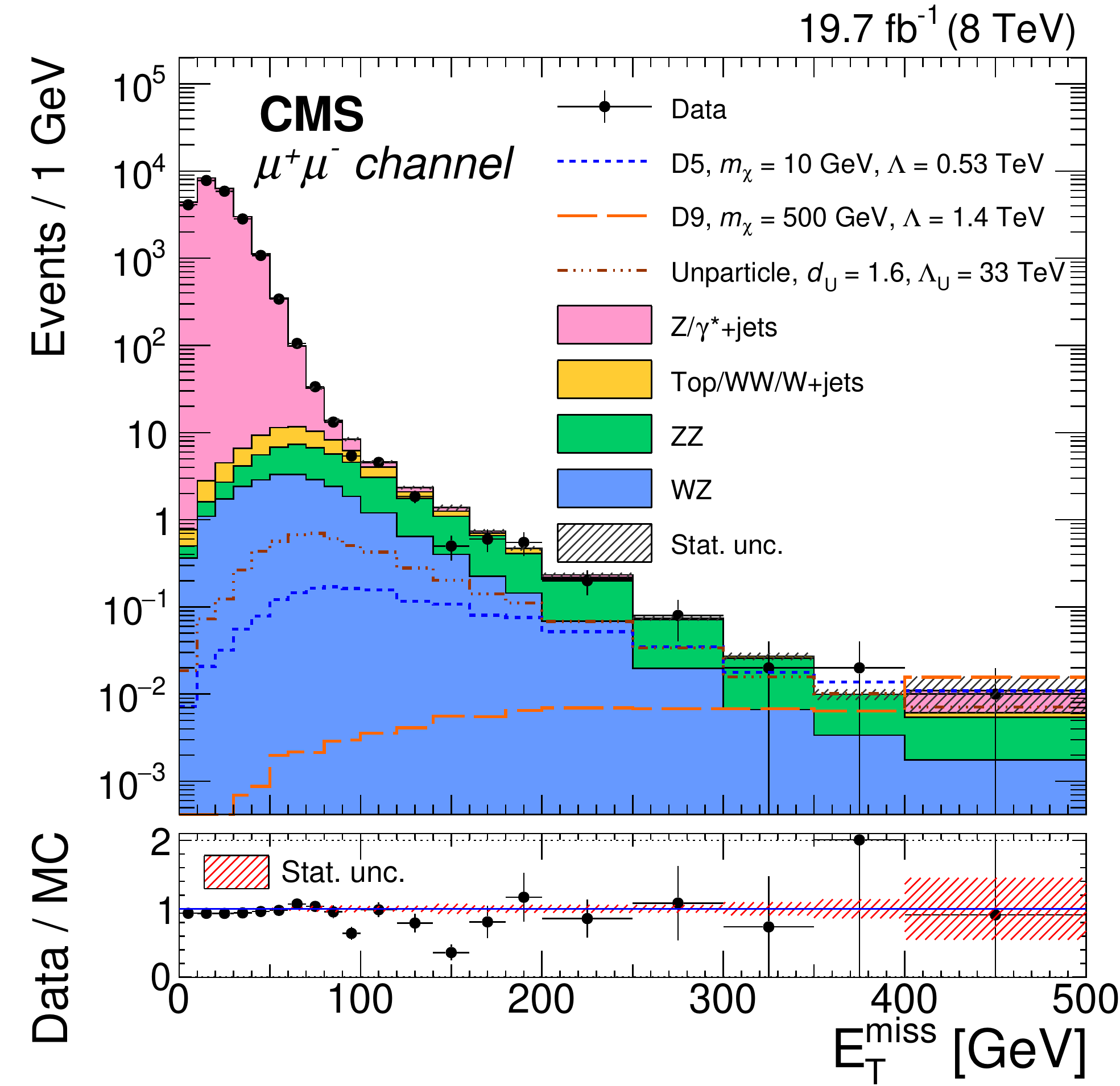}
\caption{The distribution of \ETm after preselection for the $\Z \to \Pep\Pem$ (\cmsLeft) and $\Z \to \Pgmp\Pgmm$ (\cmsRight) channels.
	 Expected signal distributions are shown for Dirac fermions with vector or tensor couplings and for unparticles.
	 The total statistical uncertainty in the overall background is shown as a hatched region.
	 The horizontal bars on the data points indicate the bin width.
	 Overflow events are included in the rightmost bins.
	}
\label{fig:pfmet_presel}
\end{figure}

The final selection is optimized for DM and unparticle signals to obtain
the best expected cross section limit at 95\% \CL using four variables, \ETm,
$\Delta \phi_{\ell\ell,\ptvecmiss}$, ${\abs{\ETm-\pt^{\ell\ell}}/\pt^{\ell\ell}}$, and $u_{\parallel}/\pt^{\ell\ell}$,
where $u_{\parallel}$ is defined as the component of
$\vec{u} =-\ptvecmiss-\vec{\pt}^{\ell\ell}$
parallel to the direction of $\vec{\pt}^{\ell\ell}$.
The last three variables effectively suppress background processes such as DY and top-quark production.
If the best expected significance is used in the optimization,
instead of the best expected limit, very similar results are obtained.
In both electron and muon channels, a mass-independent event selection followed by a fit to the shape of the
transverse mass $\mt = \sqrt{\smash[b]{2 \pt^{\ell\ell} \ETm (1-\cos \Delta \phi_{\ell\ell,\ptvecmiss})}}$
distribution is used to discriminate between the signal and the backgrounds.
For each set of selection requirements considered, the full
analysis, including the estimation of backgrounds
and the systematic uncertainties, is repeated.
The final selection criteria obtained after optimization for both the electron and muon channels are
$\ETm > 80\GeV$, $\Delta \phi_{\ell\ell,\ptvecmiss} > 2.7 $, $\abs{u_{\parallel}/\pt^{\ell\ell}}<1$, and
$\abs{\ETm-\pt^{\ell\ell}}/\pt^{\ell\ell} < 0.2$.
This common selection is applied to both the DM and unparticle searches because the optimization results are very similar for both signals.
A summary of the preselection and final selection criteria for the final analysis is listed in Table~\ref{tab:selectioncuts}.
Figure~\ref{fig:pfmet_presel} shows the distributions of \ETm after preselection, in the $\Pe\Pe$ and $\Pgm\Pgm$ channels.
Good agreement is found between the observed distributions and the background
prediction, which is described in the following section.

\begin{table}[!htb]
  \centering
  \topcaption{Summary of selections used in the analysis.}
  \begin{scotch} {llll}
                                & Variable & \multicolumn{2}{l}{Requirements}\\
  \cline{2-4}\noalign{\vskip 1mm}
\multirow{5}{*}{Preselection}   & $\pt^{\ell}$                              &   \multicolumn{2}{l}{$>$20\GeV}  \\
                                & $\abs{\mll - m_{\Z}}$                         &   \multicolumn{2}{l}{$<$10\GeV}  \\
                                & Jet counting                              &    \multicolumn{2}{l}{$\leq$1 jets with $\pt^\mathrm{j} > 30\GeV$}  \\
                                & $\pt^{\ell\ell}$                          &    \multicolumn{2}{l}{$>$50\GeV}       \\
                                & 3rd-lepton veto                           &    \multicolumn{2}{l}{$\pt^{\ell}>10$\GeV}       \\
                                & Top quark veto                            &    \multicolumn{2}{l}{veto on $\cPqb$ jets and soft muon} \\ \hline\noalign{\vskip 1mm}
\multirow{4}{*}{Selection}      & $|u_{\parallel}/\pt^{\ell\ell}|$          &    \multicolumn{2}{l}{$<$1}  \\
                                & \ETm                                    &    \multicolumn{2}{l}{$>$80\GeV}  \\
                                & $\Delta \phi_{\ell\ell,\ptvecmiss}$          &    \multicolumn{2}{l}{$>$2.7 rad}   \\
                                & $|\ETm-\pt^{\ell\ell}|/\pt^{\ell\ell}$    &    \multicolumn{2}{l}{$<$0.2}\\
  \end{scotch}
  \label{tab:selectioncuts}
 \end{table}

\section{Background estimation}
\label{sec:backgrounds}
The $\ZZ$ and $\WZ$ backgrounds are modeled using MC simulation, and
normalized to their respective NLO cross sections computed with \MCFM6.8~\cite{MCFM}. Other backgrounds, including
$\ttbar$, $\tw$, $\WW$, $\Z \to\Pgt\Pgt$, and DY are estimated from data for the final selection.
The background from \Wjets is negligible in the muon channel but significant in the
electron channel, where an estimation method based on control samples in data is used for its estimation.

The background processes that do not involve $\Z$ boson production are referred to as nonresonant
backgrounds. Such backgrounds arise mainly from leptonic $\W$ boson decays in $\ttbar$, $\tw$,
and $\WW$ events.
There are also small contributions from $s$- and $t$-channel single top quark events
and $\Z \to\Pgt\Pgt$ events in which $\Pgt$ leptons produce
electrons or muons and \ETm.
We estimate these nonresonant backgrounds using a data control sample, consisting of events with
an opposite-charge different-flavor dilepton pair ($\Pe^{\pm}\Pgm^{\mp}$) that otherwise pass
the full selection.
As the decay rates for $\Z \to\Pep\Pem$ and $\Z \to\Pgmp\Pgmm$ are equal, by equating
the ratio of observed dilepton counts to the square of the ratio of
efficiencies,
the backgrounds in the $\Pe\Pe$ and $\Pgm\Pgm$ channels can be estimated,
\begin{align*}
N^{\text{est}}_{\text{bkg},\Pe\Pe} &= N^{\text{data, corr}}_{\Pe\Pgm} \, k_{\Pe\Pe}, &
k_{\Pe\Pe} &= \frac12 \sqrt{\frac{N^{\text{data}}_{\Pe\Pe}}{N^{\text{data}}_{\Pgm\Pgm}}},\\
N^{\text{est}}_{\text{bkg},\Pgm\Pgm} &= N^{\text{data, corr}}_{\Pe\Pgm} \, k_{\Pgm\Pgm}, &
k_{\Pgm\Pgm} &= \frac12 \sqrt{\frac{N^{\text{data}}_{\Pgm\Pgm}}{N^{\text{data}}_{\Pe\Pe}}},
\end{align*}
in which the coefficient of $1/2$ in the correction factors $k_{\Pe\Pe}$ and $k_{\Pgm\Pgm}$
comes from the dilepton decay ratios for $\Pe\Pe$, $\Pgm\Pgm$, and $\Pe\Pgm$ in these nonresonant backgrounds
and $N^\text{data}_{\Pe\Pe}$ and $N^\text{data}_{\Pgm\Pgm}$ are the numbers of selected $\Pe\Pe$ and $\Pgm\Pgm$ events
from data with masses inside the $\Z$ mass window.
The ratio $\sqrt{\smash[b]{N^\text{data}_{\Pe\Pe}/N^\text{data}_{\Pgm\Pgm}}}$ and
the reciprocal quantity take into account the difference between the electron and muon selection efficiencies.
The term $N^\text{data, corr}_{\Pe\Pgm}$ is the number of $\Pe\Pgm$ events
observed in data corrected by subtracting $\ZZ, \WZ$, DY,
and \Wjets background contributions estimated using MC simulation.
The kinematic distributions of the estimated nonresonant backgrounds are taken from the distributions of the $\Pe\Pgm$ sample with the overall normalization determined by the method described above.
The validity of this procedure for predicting nonresonant backgrounds is checked with simulated events containing
$\ttbar$, $\tw$, $\WW$, and $\Z \to\Pgt\Pgt$ processes.
We assign a systematic uncertainty of 17\% (15\%) to this background estimation
in the electron (muon) channel, based on closure tests that compare the predictions obtained from the control sample with those from the simulated events.

The DY process is dominant in the region of low \ETm.
This process does not produce undetectable particles, and therefore the measured \ETm arises from
limited detector acceptance and mismeasurement.
The estimation of this background uses simulated DY events, which are normalized to the data with scale factors
obtained by measuring the number of DY events in background-dominated control regions,
after subtracting other processes. These scale factors are of order 1.1--1.2.
The control regions are defined with the full selection except for the requirements
on \ETm, $\Delta \phi_{\ell\ell,\ptvecmiss}$, and $\abs{\ETm-\pt^{\ell\ell}}/\pt^{\ell\ell}$.
The results are calculated independently for control regions with variables \ETm and $\abs{\ETm-\pt^{\ell\ell}}/\pt^{\ell\ell}$ and
compared with each other as part of the estimate of systematic uncertainty.
Based on the variations of the estimates with the choice of control regions,
a systematic uncertainty of 10\% (11\%) is assigned to the DY background estimate in the electron (muon) channel.

A \Wjets background event consists of a genuine prompt lepton from the \W decay and a nonisolated
lepton resulting from the leptonic decay of heavy quarks, misidentified hadrons, or electrons from
photon conversions. The rate at which jets are misidentified as leptons may not be accurately
described in the MC simulation, so the rate of
jets passing lepton identification requirements is determined using a control data sample enriched in jets.
The genuine lepton contamination from \W/\Zjets
events in the selected control sample is subtracted using simulation to avoid biasing the calculation of the misidentification rate.
The final estimation is obtained by applying these weights to a sample selected
with lepton identification requirements that are looser than for the signal sample.
The main source of systematic uncertainty for this background estimation
comes from the measurement of the misidentification rate, which depends on the accuracy of the subtraction of real leptons in the control sample.
A systematic uncertainty of 15\% is assigned, based on the dependence of the calculated misidentification rates
on the selection criteria applied to the control sample.

\section{Efficiencies and systematic uncertainties}
\label{sec:systematics}
The efficiencies for selecting, reconstructing, and identifying isolated leptons are determined
from simulation and then corrected with scale factors determined from applying a ``tag-and-probe'' technique~\cite{CMS:2011aa} to $\Z \to\ell^+\ell^-$ events.
The trigger efficiencies for the electron and muon
channels are found to be above 90\%, varying as a function of $\pt$ and $\abs{\eta}$ of the lepton. The identification
efficiency for electrons (muons), when applying the criteria described in Sec.~\ref{sec:object}, is found to be 95\% (94\%).
The corresponding data-to-MC scale factors are typically in the range 0.94--1.01 (0.98--1.02) for the electron (muon) channel,
depending on the $\pt$ and $\abs{\eta}$ of the lepton candidate.
For both channels, the overall uncertainty in selecting and reconstructing leptons in an event is about 3\%.

The systematic uncertainties include normalization uncertainties that affect the overall size of contributions, and shape
uncertainties that alter the shapes of the distributions used in extracting the signal limits. The systematic
uncertainties are summarized in Table~\ref{tab:syst}.

\begin{table*}[htb]
\centering
\topcaption{Summary of systematic uncertainties. Each background uncertainty represents the variation of the relative yields of the particular background
components. The signal uncertainties represent the relative variations in the signal acceptance, and
ranges quoted cover both signals of DM and unparticles with different DM masses or scaling dimensions.
For shape uncertainties, the numbers correspond to
the overall effect of the shape variation on yield or acceptance.
The symbol \NA\ indicates that the systematic uncertainty is not applicable.
}
\begin{scotch}{lccc}
Source & Background & Signal \\
  		    & uncertainty (\%) & uncertainty (\%) \\
\hline
  PDF$+\alpha_S$                            & 5--6 & 8--20  \\
  Factorization and renormalization scale       & 7--8 & 5  \\
  Acceptance ($\ZZ$)	                    & 14   &\NA\\
  Integrated luminosity                     & 2.6  & 2.6  \\
  Lepton trigger, reconstruction \& identification, isolation     & 3    & 3  \\
  DY normalization                     & 10--11 &\NA\\
  $\ttbar$, $\tw$, $\WW$ normalization      & 15--17    &\NA\\
  $\Wjets$ normalization                    & 15--23 &\NA\\
  MC statistics (signal, $\ZZ$, $\WZ$)                   & 1--2 & 1--2 \\
  Control region statistics (DY)       		 & 25  &\NA\\
  Control region statistics ($\ttbar$, $\tw$, $\WW$)     & 18  &\NA\\
  Control region statistics ($\Wjets$)                   & 36  &\NA\\
  Pileup                              		         & 0.5--1   & 0.1--0.7 \\
  $\cPqb$-jet tagging efficiency                         & 0.4--1.4 & 0.6--1 \\
  Lepton momentum scale                		         & 0.4--0.5 & 0.1--1 \\
  Jet energy scale and resolution      		         & 5--7     & 3--5 \\
  Unclustered \ETm\ scale                                & 1--2     & 1 \\
\end{scotch}
\label{tab:syst}
\end{table*}

The normalization uncertainties in the background estimates from data are described
in Sec.~\ref{sec:backgrounds}.
The overall approach for the estimation of the PDF and $\alpha_{S}$ uncertainties
(referred to as PDF$+\alpha_{S}$ in the following) adopts the interim recommendations
of the PDF4LHC group and is used both for signal and the
background~\cite{Alekhin:2011sk,Botje:2011sn,Ball:2012cx,Nadolsky:2008zw,Martin:2009iq}.
This is the most important uncertainty for the signals.
As the mass of the DM particles increases, the PDF$+\alpha_{S}$ uncertainty reaches 20\%, which can be explained by the
diminishing phase space for DM production and the rise of the corresponding uncertainty in the cross section.
The efficiencies for signal, $\ZZ$, and $\WZ$ processes are estimated using simulation, and the uncertainties
in the corresponding yields are derived from variations of the renormalization and factorization scales, $\alpha_{S}$, and choice of PDFs,
in which the factorization and renormalization scales are assessed by varying the original scales of the process by factors of 0.5 and 2.
Typical values for the signal extraction efficiency are found to be around 40\%.
The uncertainty related to the renormalization and factorization scales is 5\% for signal and 7\%--8\% for $\ZZ$ and $\WZ$ processes.
The effect of variations in $\alpha_{S}$ and the choice of PDFs is 5\%--6\% for the $\ZZ$ and $\WZ$ backgrounds.
The uncertainty assigned to the luminosity measurement is 2.6\%~\cite{CMS-PAS-LUM-13-001}.

The contributions to the shape uncertainties come from the lepton momentum scale, the jet energy scale and resolution,
the unclustered \ETm scale, the $\cPqb$ tagging efficiency, and the pileup modeling.
Each corresponding uncertainty is calculated by varying the respective variable of interest within its
own uncertainties and propagating the variations to the variable $\mt$ using the final selection.
In the case of the lepton momentum scale, the uncertainty is computed by varying the momentum of the leptons by their uncertainties.
The uncertainty in the muon momentum scale is 1\%.
For electrons, uncertainties of 0.6\% for the barrel and 1.5\% for the end caps are applied.
For the $\ZZ$ background, cross checks are made using the generators \MADGRAPH, \POWHEG~\cite{Nason:2013ydw},
and \SHERPA2.1.1~\cite{Gleisberg:2008ta}. A comparison of the acceptance from normalized yields obtained with these generators
is made in the signal region.
In each bin of $\mt$, the maximum difference in acceptance with respect to the \MADGRAPH prediction is assigned
as a separate additional systematic uncertainty.
The limit setting procedure discussed in Sec.~\ref{sec:results} incorporates this acceptance uncertainty
from each of the separate bins. The weighted average of these uncertainties is 14\%.
This is the dominant uncertainty in the total background prediction for the signal region. For the $\WZ$ background, this difference in acceptance is not
observed, and data and simulation agree in the selected three-lepton control region.

The uncertainties in the calibration of the jet energy scale and resolution directly affect
the assignments of jets to jet categories,
the \ETm computation, and all the selections related to jets. The
effect of the jet energy scale uncertainty is estimated by varying the energy scale by $\pm1\sigma$.
A similar strategy is used to evaluate the systematic uncertainty related to the jet energy resolution.
The uncertainties in the final yields are found to be 3\%--5\% (5\%--7\%) for signal (background).
The effect of the uncertainty in the energy scale of the unclustered component
of the \ETm measurement is estimated by subtracting the leptons and jets from
the \ETm summation and by varying the residual recoil by $\pm$10\%.
The clustered component is then added back in order to recalculate the value of \ETm.
The resultant uncertainty in the final yields is found to be of order 1\%--2\%.
Since the $\cPqb$ tagging efficiencies measured in data are somewhat different from those
predicted by the simulation, an event-by-event reweighting using data-to-MC scale factors
is applied to simulated events.
The uncertainty associated with this procedure is obtained by varying the event-by-event weight by ${\pm}1\sigma$.
The total uncertainty in the final yields is 0.6--1\% (0.4--1.4\%) for signal (background).
All simulated events are reweighted to reproduce the pileup conditions observed in data.
To compute the uncertainty related to pileup modeling, we shift the mean of the distribution in simulation by 5\%.
The variation of the final yields induced by this procedure is less than 1\%.
For the processes estimated from simulation, the sizes of the MC samples limit the precision of the modeling, and
the corresponding statistical uncertainty is incorporated into the shape uncertainty. A similar treatment is applied to the backgrounds
estimated from control samples in data based on the statistical uncertainties in the corresponding control samples.

\section{Results}
\label{sec:results}
\begin{figure}[htb!]
\centering
\includegraphics[width=0.48\textwidth]{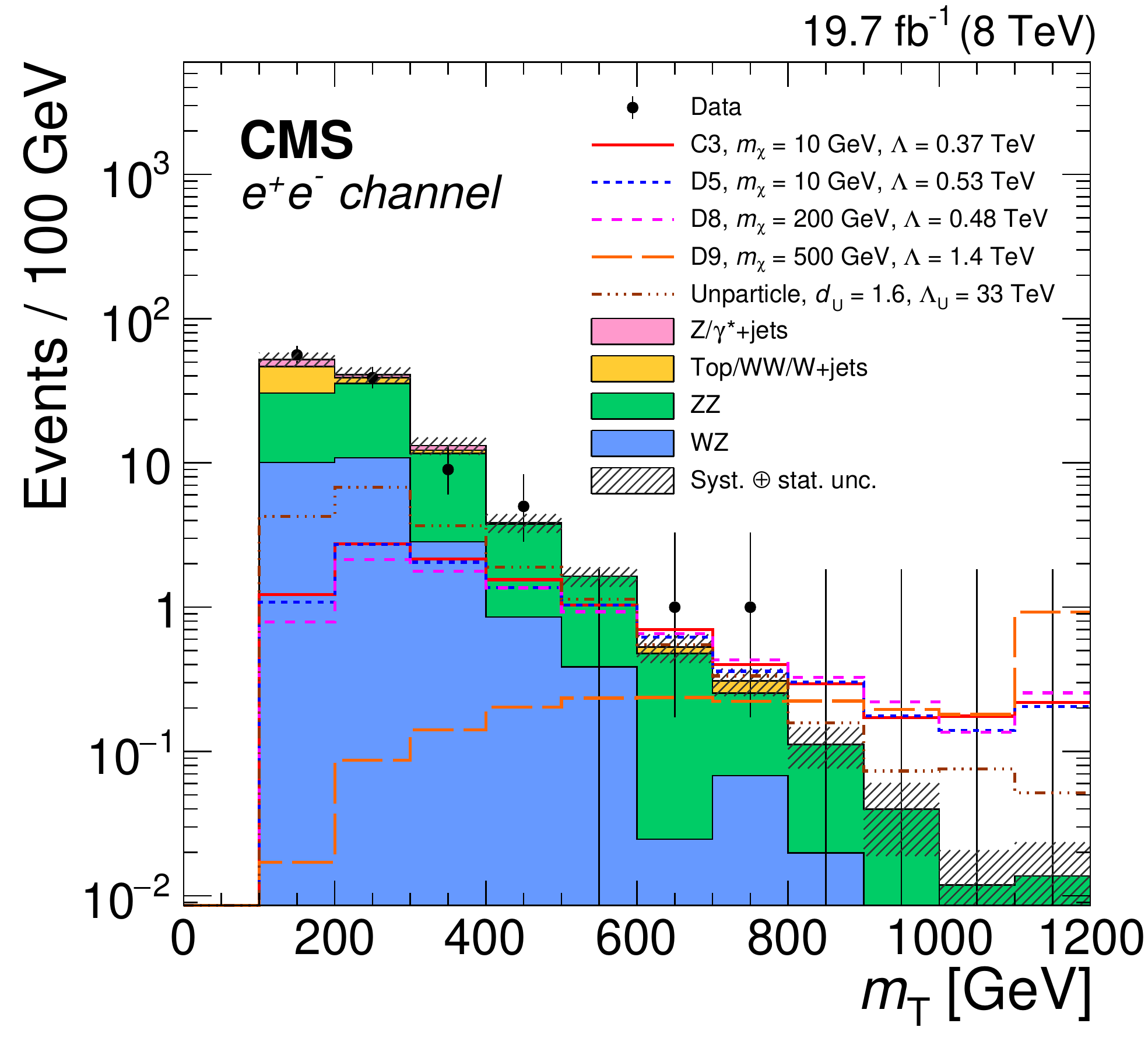}
\includegraphics[width=0.48\textwidth]{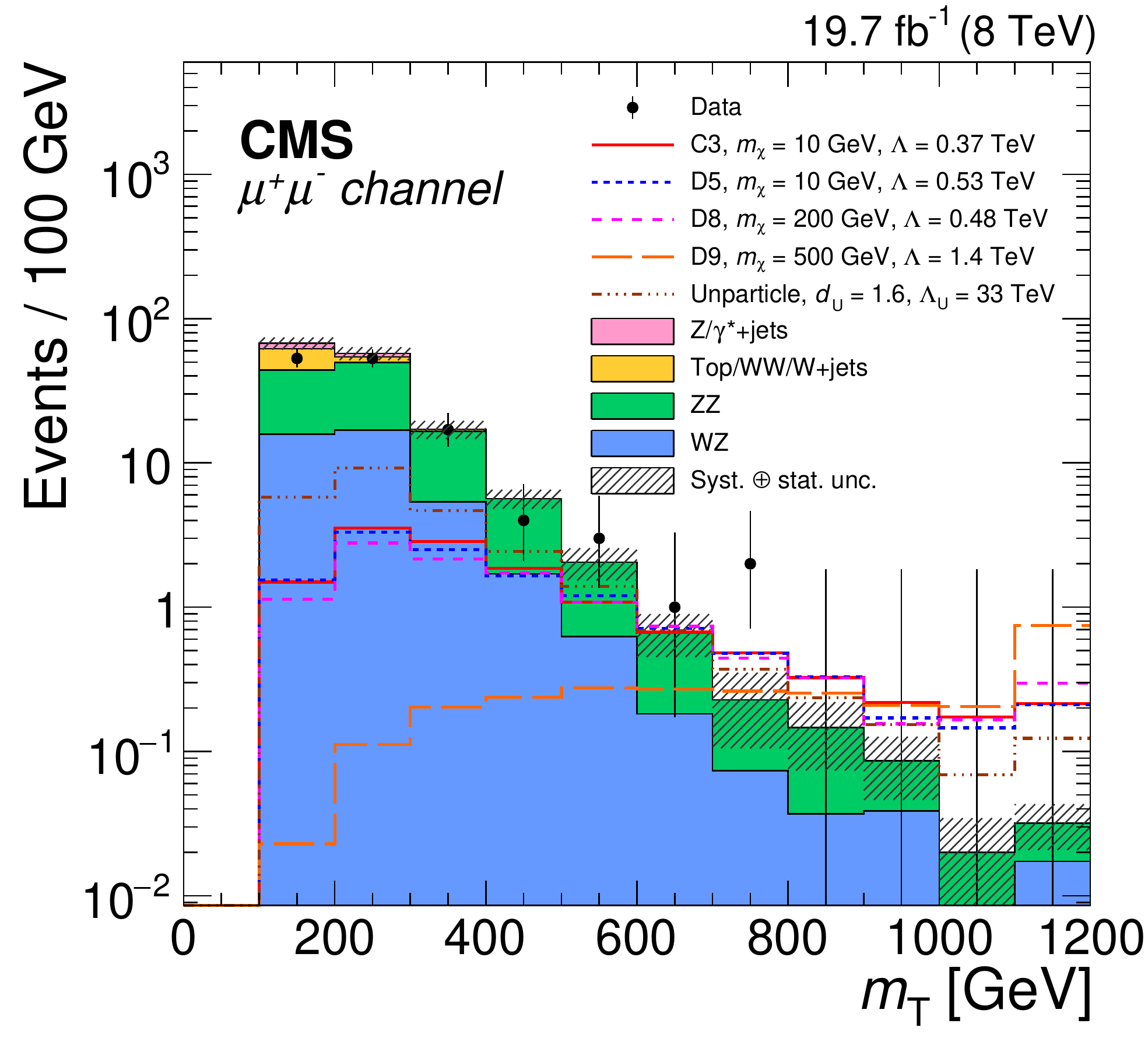}
\caption{Distributions of the transverse mass
	for the final selection in the $\Pep\Pem$ (\cmsLeft) and $\Pgmp\Pgmm$ (\cmsRight) channels.
	Examples of expected signal distributions are shown for DM particle production and unparticle production.
	The total statistical and systematic uncertainty in the
	overall background is shown as a hatched region.
	Overflow events are included in the rightmost bins.
        }
\label{fig:postfit_shapes_LogY}
\end{figure}

For both the electron and the muon channels, a shape-based analysis is employed.
The expected numbers of background and signal events scaled by a signal strength modifier
are combined in a binned likelihood for each bin of the $\mt$ distribution.
The signal strength modifier, defined as the signal cross section divided by
the cross section suggested by theory, determines the strength of the signal process.
The numbers of observed and expected events are shown in Table~\ref{tab:seltable},
including the expectation for a selected mass point for each type of signal.
Figure~\ref{fig:postfit_shapes_LogY} shows the $\mt$ distributions after the final selection.
The observed distributions agree with the SM background predictions and no excess of events is observed.

\begin{table*}[h!tb]
\centering
\topcaption{Signal predictions, background estimates, and observed number of events.
	The DM signal yields are given for masses $m_\chi=$ 10, 200, and 500\GeV
	and cutoff scales $\Lambda=0.37$, 0.53, 0.48, and $1.4\TeV$.
	The yields from an unparticle signal are presented with a scaling dimension $d_\mathcal{U}=1.6$
	and a renormalization scale $\Lambda_\mathcal{U}=33\TeV$.
	The corresponding statistical and systematic uncertainties are shown, in that order.}
\begin{scotch}{lll}
Process & $\Pep\Pem$ & $\Pgmp\Pgmm$ \\
\hline
C3, $m_\chi=10\GeV$, $\Lambda = 0.37\TeV$       		& 10.7	$\pm$	0.2	$\pm$	1.1	&	12.8	$\pm$	0.3	$\pm$	1.1 \\
D5, $m_\chi=10\GeV$, $\Lambda = 0.53\TeV$       		& 10.0	$\pm$	0.3	$\pm$	1.1	&	12.3	$\pm$	0.3	$\pm$	1.1 \\
D8, $m_\chi=200\GeV$, $\Lambda = 0.48\TeV$      		& 9.0	$\pm$	0.2	$\pm$	1.1	&	11.1	$\pm$	0.2	$\pm$	0.9 \\
D9, $m_\chi=500\GeV$, $\Lambda = 1.4\TeV$       		& 2.67	$\pm$	0.03	$\pm$	0.41	&	2.81	$\pm$	0.03	$\pm$	0.26 \\
Unparticle, $d_\mathcal{U}=1.6$, $\Lambda_\mathcal{U}=33\TeV$ 	& 19.0	$\pm$	0.3	$\pm$	1.3	&	25.6	$\pm$	0.4	$\pm$	1.7 \\
\hline
     				  $\dyll$ & 8.2 $\pm$ 1.9 $\pm$ 0.8	 & 8.6 $\pm$ 3.0 $\pm$ 1.0  \\
                $\WZ\to 3\ell\nu$ & 25.1 $\pm$ 0.5 $\pm$ 2.8     & 40.7 $\pm$ 0.7 $\pm$ 4.5 \\
               $\ZZ\to 2\ell2\nu$ & 59 $\pm$ 1 $\pm$ 10          & 79 $\pm$ 1 $\pm$ 14 \\
    $\ttbar$/$\tw$/\WW/$\Z \to\Pgt\Pgt$ 	  & 18.7 $\pm$ 3.4 $\pm$ 3.3     & 22.9 $\pm$ 2.3 $\pm$ 3.4 \\
                                   \Wjets & 1.8 $\pm$ 0.6 $\pm$ 0.3      & \NA\\
\hline
			 Total background & 113 $\pm$ 4 $\pm$ 13    & 151 $\pm$ 4 $\pm$ 18 \\
\hline
                                    Data & 111   & 133 \\
\end{scotch}
\label{tab:seltable}
\end{table*}

Upper limits on the contribution of events from new physics are computed by
using the modified frequentist approach CL$_\mathrm{s}$~\cite{Read1,junkcls} based on asymptotic
formulas~\cite{Cowan:2010js,HiggsCombination}.

\subsection{DM interpretation}

\begin{figure*}[htb!]
\centering
\includegraphics[width=0.48\textwidth]{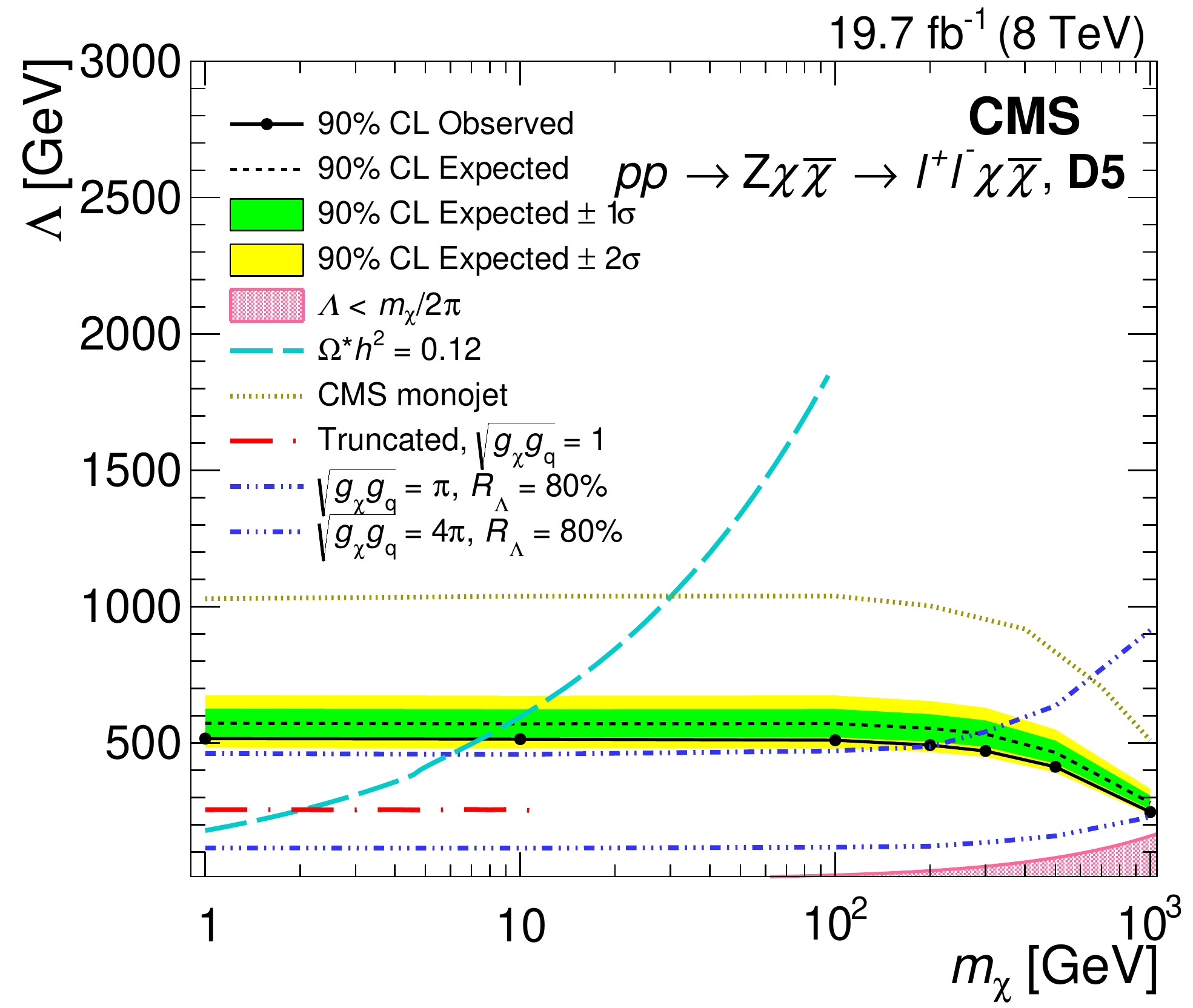}
\includegraphics[width=0.48\textwidth]{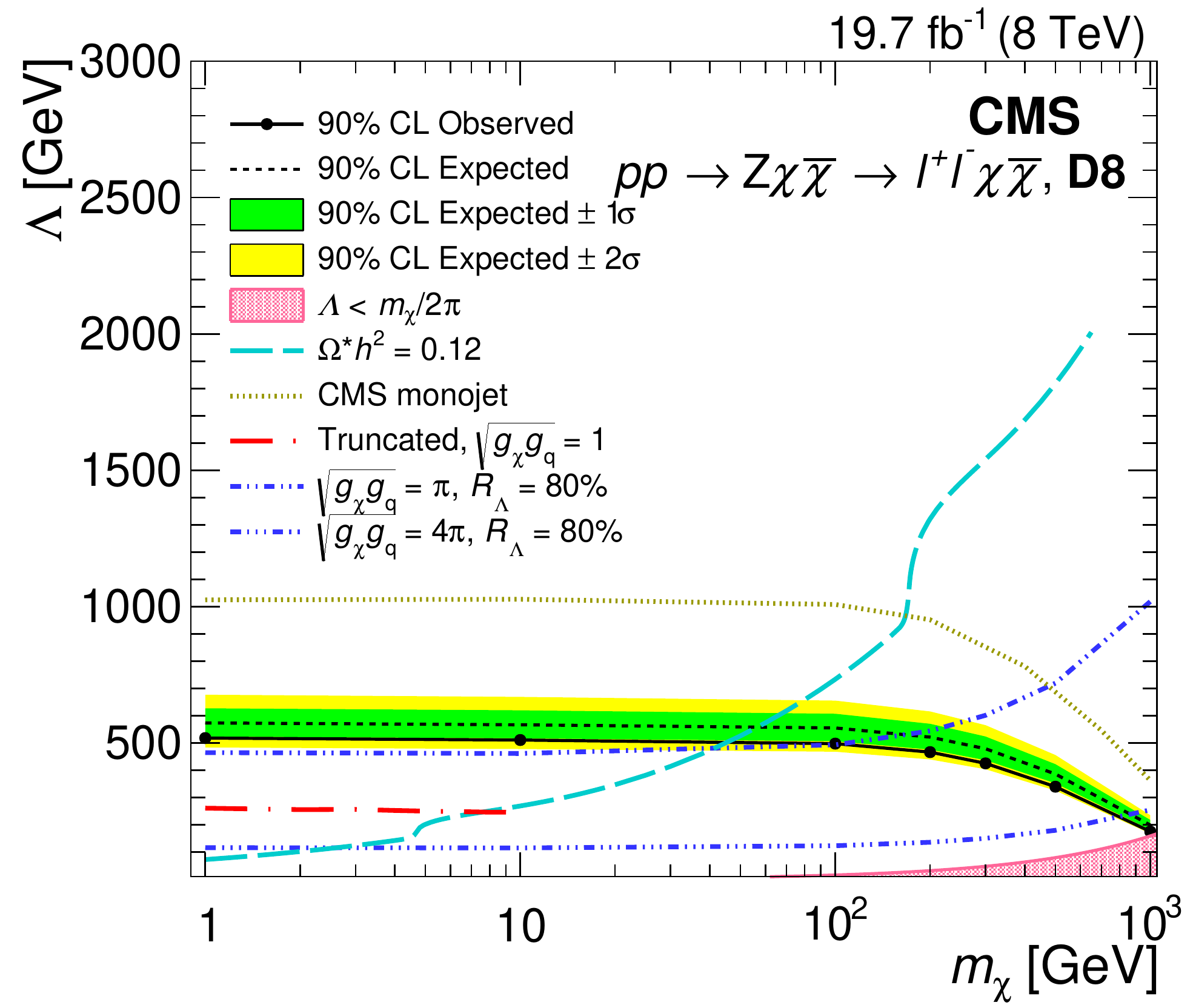}
\includegraphics[width=0.48\textwidth]{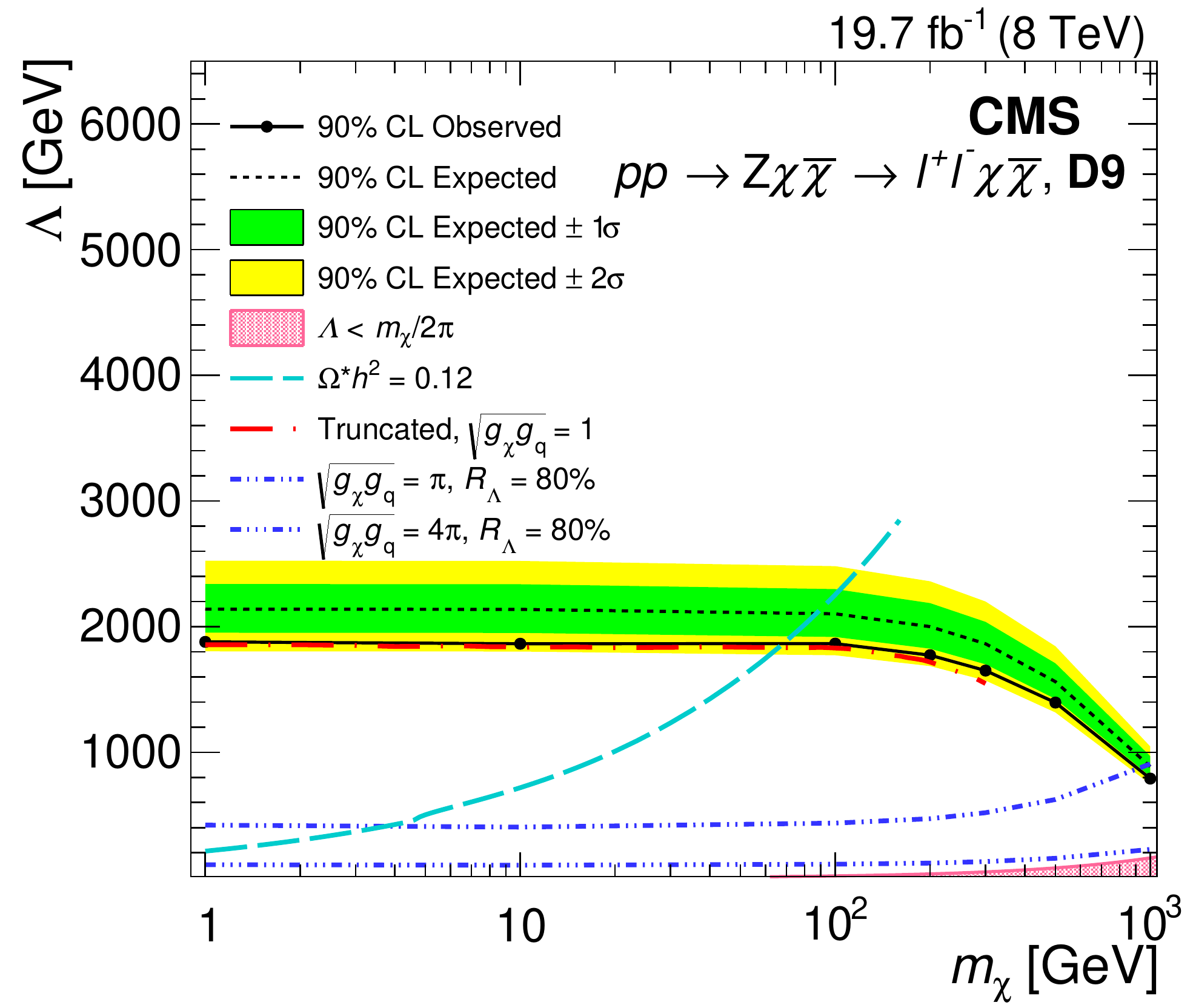}
\includegraphics[width=0.48\textwidth]{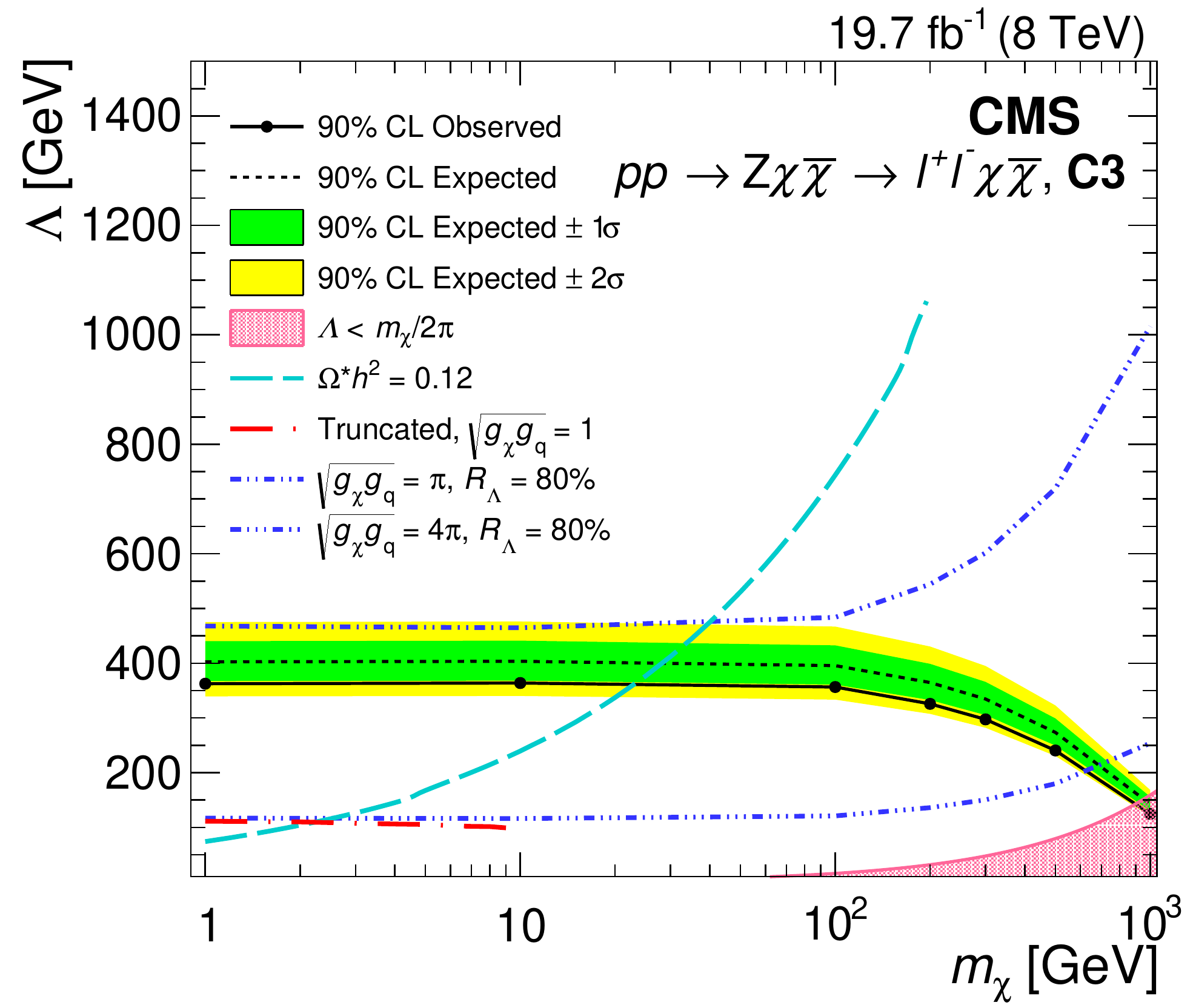}
\caption{Expected and observed 90\% \CL lower limits on $\Lambda$ as a function of DM particle mass $m_{\chi}$
        for the operators D5 (top left), D8 (top right), D9 (bottom left), and C3 (bottom left).
	The pink shaded area is shown in each plot to indicate the lower
	bound $\Lambda > m_{\chi}/2\pi$ on the validity of the effective field theory DM model.
	The cyan long-dashed line calculated by MadDM~1.0~\cite{Backovic:2013dpa} reflects the relic density of cold, nonbaryonic
	DM: $\Omega h^2=0.1198\pm0.0026$ measured by the Planck telescope~\cite{Ade:2013zuv}.
	Monojet results from CMS~\cite{Khachatryan:2014rra} are shown for comparison.
        Truncated limits with $\sqrt{g_\cPq g_\chi}=1$ are presented with red dot long-dashed lines.
	The blue double-dot and triple-dot dashed lines indicate the contours of $R_{\Lambda}=80\%$ for all operators with couplings $\sqrt{g_\cPq g_\chi}=\pi$ and $4\pi$.
	}
\label{fig:DM_Lambda_Limits}
\end{figure*}

The observed limit on the cross section for DM production depends on the DM particle mass and
the nature of DM interactions with SM particles. Within the framework of effective field
theory, the upper limits on this cross section can be translated into 90\% \CL lower limits on
the effective cutoff scale $\Lambda$ as a function of DM particle mass $m_{\chi}$, as shown in
Fig.~\ref{fig:DM_Lambda_Limits}.
The choice of 90\% \CL is made in order to allow comparisons with direct detection experiments.
The relic density of cold, nonbaryonic DM has been measured by Planck telescope~\cite{Ade:2013zuv}
using the anisotropy of the cosmic microwave background and of the spatial distribution of galaxies.
They obtain a value $\Omega h^2=0.1198\pm0.0026$, where $h$ is the Hubble constant.
The implications of this result plotted in the plane of the effective cutoff scale $\Lambda$ and DM mass
$m_{\chi}$ have been calculated with MadDM~1.0~\cite{Backovic:2013dpa}, and are shown in Fig.~\ref{fig:DM_Lambda_Limits}.
Results from a search for DM particles using monojet signatures in CMS~\cite{Khachatryan:2014rra} are also plotted for comparison.

It has been emphasized by several authors~\cite{Goodman:2010ku,TevatronDMFrontier,Friedland:2011za,Buchmueller:2013dya}
that the effective field theory approach is not valid over the full range of phase space that is accessible at the LHC,
since the scales involved can be comparable to the collision energy. In the LHC regime, the assumption of a pointlike interaction
provides a reliable approximation of the underlying ultraviolet-complete theory only for appropriate choices of couplings and masses.
To estimate the region of validity relevant to this analysis, we consider a simple tree-level ultraviolet-complete model that contains
a massive mediator ($M$) exchanged in the $s$ channel, with the couplings to quarks and DM particles described by coupling constants $g_\cPq$ and $g_\chi$.
The effective cutoff scale $\Lambda$ thus can be expressed as $\Lambda \sim M/\sqrt{g_\cPq g_\chi}$, when momentum transfer is small ($Q_\text{tr}<M$).
Imposing a condition on the couplings $\sqrt{g_\cPq g_\chi}<4\pi$ to ensure stability of the perturbative calculation and a mass requirement $M>2m_\chi$,
a lower bound $\Lambda > m_\chi/2\pi$ is obtained for the region of validity. The area below this boundary, where the effective theory of DM
is not expected to provide a reliable prediction at the LHC, is shown as a pink shaded area in each of the panels of Fig.~\ref{fig:DM_Lambda_Limits}.

However, the requirement of $\Lambda > m_\chi/2\pi$ is not sufficient, according to some authors~\cite{TevatronDMFrontier,
Fox:2011pm,Goodman:2011jq,Shoemaker:2011vi,Buchmueller:2013dya,Busoni:2013lha,Busoni:2014sya,Busoni:2014haa,
Malik:2014ggr,Abdallah:2015ter,Abercrombie:2015wmb},
and the region of validity depends on the coupling values in the ultraviolet completion of the theory.
Considering a more realistic minimum constraint $Q_\text{tr} < M \sim \sqrt{g_\cPq g_\chi} \Lambda$,
we can calculate the ratio $R_{\Lambda}$ of the number of events fulfilling the validity criteria over all events
produced in the accessible phase space,
\begin{equation*}
R_{\Lambda} = \frac{\left.\int_{\pt^\text{min}}^{\pt^\text{max}}\rd\pt \int_{\eta^\text{min}}^{\eta^\text{max}}
	\rd\eta \frac{\rd^2\sigma_\text{eff}}{\rd\pt \rd\eta}\right|_{Q_\text{tr} <\sqrt{g_\cPq g_\chi} \Lambda}}
{\int_{\pt^\text{min}}^{\pt^\text{max}}\rd\pt \int_{\eta^\text{min}}^{\eta^\text{max}}\rd\eta \frac{\rd^2\sigma_\text{eff}}{\rd\pt \rd\eta} },
\end{equation*}
in which the values of $R_{\Lambda}$ can be used to check the accuracy of the effective description in regions of parameter space $(\Lambda,m_\chi)$.
Figure~\ref{fig:DM_Lambda_Limits} includes the corresponding contours of $R_{\Lambda}=80\%$ for all operators with couplings $\sqrt{g_\cPq g_\chi}=\pi$, and $4\pi$.
Alternatively, we can obtain the truncated limits by manually removing the events with $Q_\text{tr} > \sqrt{g_\cPq g_\chi}\Lambda $ at the generator level.
Figure~\ref{fig:DM_Lambda_Limits} also shows these truncated limits with $\sqrt{g_\cPq g_\chi}=1$.
For a certain value of $m_\chi$, the truncated limit goes to zero quickly because none of the events above this value fulfills the requirement $Q_\text{tr} < \sqrt{g_\cPq g_\chi}\Lambda$.
For a maximum coupling $g_{\chi,\cPq}=4\pi$, 100\% of the events
passes this requirement, and the truncated limits coincide with the observed one and are not shown.

\begin{figure}[htb!]
\centering
\includegraphics[width=0.48\textwidth]{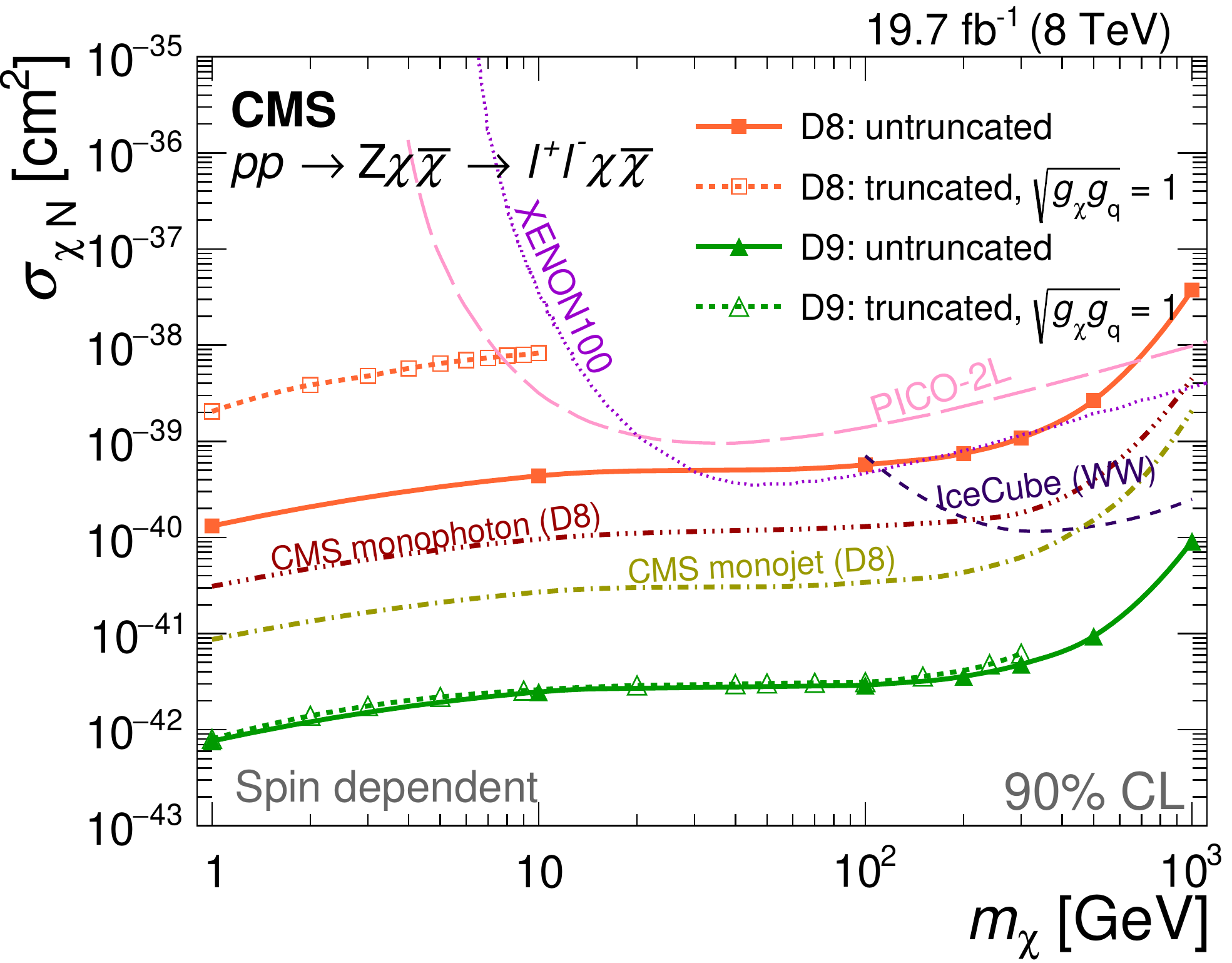}
\includegraphics[width=0.48\textwidth]{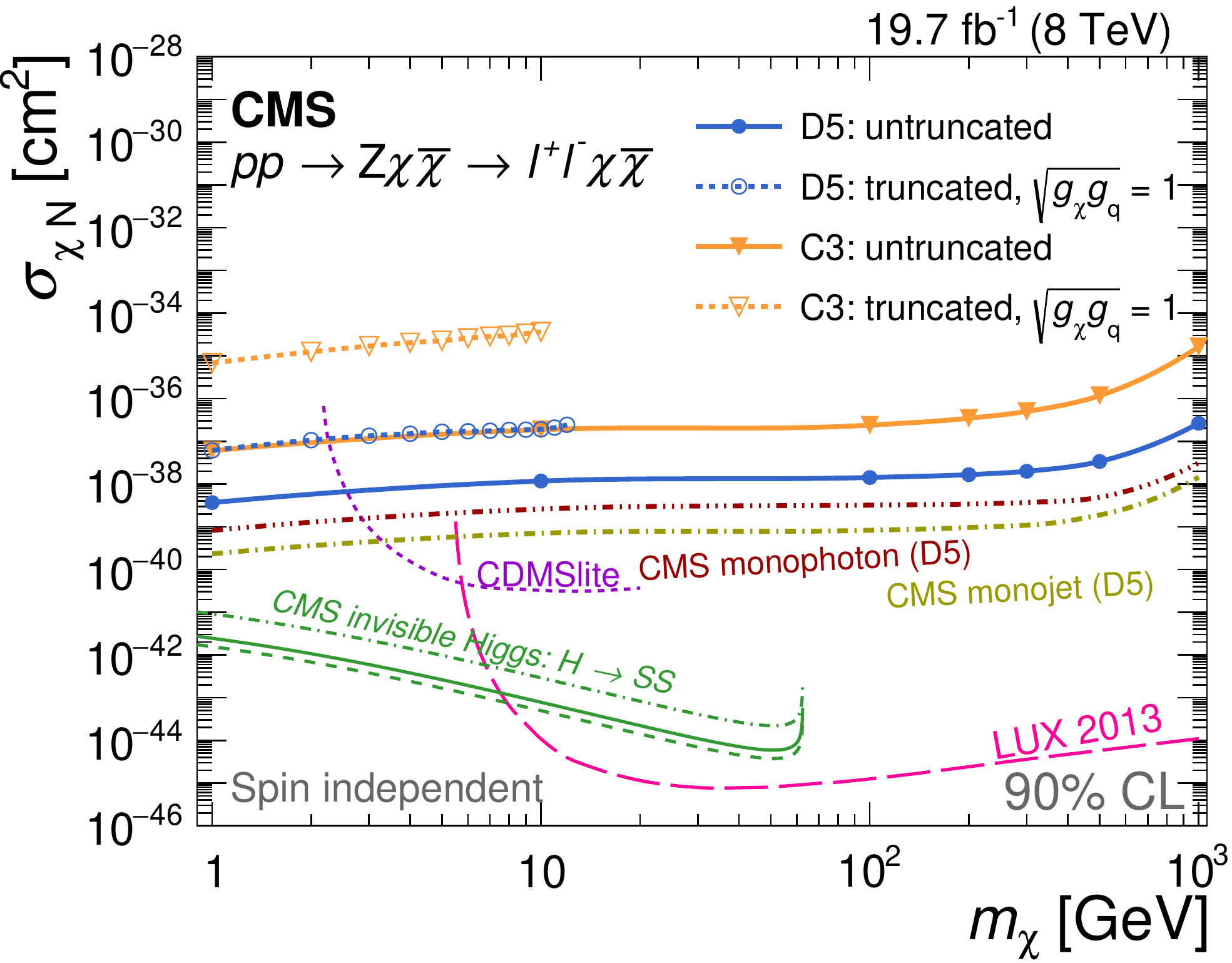}
\caption{The 90\%~\CL upper limits on the DM-nucleon cross section as a function of the DM particle mass.
	\cmsLLeft: spin-dependent limits for axial-vector (D8) and tensor (D9) coupling
	of Dirac fermion DM candidates, together with direct search experimental results
	from
	the PICO~\cite{Amole:2015lsj}, XENON100~\cite{Aprile:2013doa}, and IceCube~\cite{IceCube:2011aj} collaborations.
	\cmsRRight: spin-independent limits for vector coupling of complex scalar (C3) and
	Dirac fermion (D5) DM candidates, together with CDMSlite~\cite{Agnese:2013jaa}, LUX~\cite{Akerib:2013tjd},
	as well as Higgs-portal scalar DM results from CMS~\cite{Chatrchyan:2014tja} with central (solid), minimum (dashed) and maximum (dot dashed)
	values of Higgs-nucleon couplings.
        Collider results from CMS monojet~\cite{Khachatryan:2014rra} and monophoton~\cite{CMSMonoPhoton8TeV} searches,
	interpreted in both spin-dependent and spin-independent scenarios, are shown for comparison.
	The truncated limits for D5, D8, D9, and C3 with
	$\sqrt{g_\cPq g_\chi}=1$ are presented with dashed lines in same shade as the untruncated ones.
	}
\label{fig:DM-NucleonXS}
\end{figure}

Figure~\ref{fig:DM-NucleonXS} shows the 90\%~\CL upper limits on the DM-nucleon cross section
as a function of DM particle mass for both the spin-dependent and spin-independent
cases~\cite{TevatronDMFrontier,micromega} obtained using the relations:
\ifthenelse{\boolean{cms@external}}{
\begin{align*}
&\sigma_{0}^{\mathrm{D8,D9}} = \sum_{\cPq} \frac{3\mu_{\chi N}^2}{\pi\Lambda^4}\, \left(\Delta_\cPq^N\right)^2\\
	&\qquad= 9.18\times10^{-40}\text{cm}^2\left(\frac{\mu_{\chi N}}{1\GeV}\right)^2\left(\frac{300\GeV}{\Lambda}\right)^4,\\
&\sigma_{0}^{\mathrm{D5}} = \sum_{\cPq} \frac{\mu_{\chi N}^2}{\pi\Lambda^4}\, \left(f_\cPq^N\right)^2\\
	&\qquad= 1.38\times10^{-37}\text{cm}^2\left(\frac{\mu_{\chi N}}{1\GeV}\right)^2\left(\frac{300\GeV}{\Lambda}\right)^4,
\\
&\sigma_{0}^{\mathrm{C3}} = \sum_{\cPq} \frac{4\mu_{\chi N}^2}{\pi\Lambda^4}\, \left(f_\cPq^N\right)^2\\
	&\qquad= 5.52\times10^{-37}\text{cm}^2\left(\frac{\mu_{\chi N}}{1\GeV}\right)^2\left(\frac{300\GeV}{\Lambda}\right)^4,
\end{align*}
}{
\begin{equation*}\begin{split}
\sigma_{0}^{\mathrm{D8,D9}} &= \sum_{\cPq} \frac{3\mu_{\chi N}^2}{\pi\Lambda^4}\, \left(\Delta_\cPq^N\right)^2
        = 9.18\times10^{-40}\text{cm}^2\left(\frac{\mu_{\chi N}}{1\GeV}\right)^2\left(\frac{300\GeV}{\Lambda}\right)^4,\\
\sigma_{0}^{\mathrm{D5}}&= \sum_{\cPq} \frac{\mu_{\chi N}^2}{\pi\Lambda^4}\, \left(f_\cPq^N\right)^2
        = 1.38\times10^{-37}\text{cm}^2\left(\frac{\mu_{\chi N}}{1\GeV}\right)^2\left(\frac{300\GeV}{\Lambda}\right)^4,\\
\sigma_{0}^{\mathrm{C3}}&= \sum_{\cPq} \frac{4\mu_{\chi N}^2}{\pi\Lambda^4}\, \left(f_\cPq^N\right)^2
        = 5.52\times10^{-37}\text{cm}^2\left(\frac{\mu_{\chi N}}{1\GeV}\right)^2\left(\frac{300\GeV}{\Lambda}\right)^4,
\end{split}\end{equation*}
}
where $\mu_{\chi N}$ is the reduced mass of the DM-nucleon system, $f_\cPq^N$ characterizes the nucleon structure
($f_\cPqu^p = f_\cPqd^n = 2$ and $f_\cPqd^p = f_\cPqu^n = 1; f=0$ otherwise), and $\Delta_\cPq^N$ represents a spin-dependent form factor
($\Delta_\cPqu^p = \Delta_\cPqd^n = 0.842\pm 0.012$, $\Delta_\cPqd^p = \Delta_\cPqu^n = -0.427\pm0.013$, $\Delta_\cPqs^p = \Delta_\cPqs^n = -0.085\pm0.018$)
as specified in Ref.~\cite{micromega}.
The truncated limits for D5, D8, D9, and C3 with $\sqrt{g_\cPq g_\chi}=1$ are presented with dashed lines in the same shade
as the untruncated ones. For comparison, direct search results as well as collider results from the CMS
monojet~\cite{Khachatryan:2014rra} and monophoton~\cite{CMSMonoPhoton8TeV} studies are shown.
The recent limits from ATLAS~\cite{ATLASMonoZ8TeV} coincide very closely with the untruncated CMS
limits on D9 and D5 operators. The ATLAS curves are not shown on the figure, since they would be difficult to distinguish by eye.
Results are also shown from a search for the invisible decays of the Higgs boson~\cite{Chatrchyan:2014tja}, interpreted
in a Higgs-portal model~\cite{Djouadi:2011aa,Djouadi:2012zc}, where a Higgs boson with a mass of 125\GeV
acts as a mediator between scalar DM and SM particles. The central (solid) line corresponds to the Higgs-nucleon
coupling value (0.326) from a lattice calculation~\cite{Young:2009zb}, and the upper (dot-dashed) and lower (dashed) lines
are maximum (0.629) and minimum (0.260) values from the MILC Collaboration~\cite{Toussaint:2009pz}.

The expected and observed limits on the effective cutoff scale $\Lambda$ as a function of the DM particle mass $m_\chi$
are listed in Tables~\ref{tab:DM_Limits_D5} and~\ref{tab:DM_Limits_D8} for the operators D5 and D8.
The values for the operators D9 and C3 are listed in Tables~\ref{tab:DM_Limits_D9} and~\ref{tab:DM_Limits_C3}.
The results are also shown in terms of limits on DM-nucleon cross sections $\sigma_{\chi N}$, to allow comparison with the results from
direct searches for DM particles.

\begin{table*}[h!tb]
\centering
\caption{Expected and observed 90\% \CL upper limits on the DM-nucleon cross section $\sigma_{\chi N}$ and
        effective cutoff scale $\Lambda$ for operator D5.}
\begin{scotch}{ccccccccc}
$m_\chi$ & \multicolumn{2}{c}{Expected} & \multicolumn{2}{c}{Expected$-1\sigma$}
       & \multicolumn{2}{c}{Expected$+1\sigma$} & \multicolumn{2}{c}{Observed} \\
\hline
      & $\Lambda$ & $\sigma_{\chi N}$ & $\Lambda$ & $\sigma_{\chi N}$ & $\Lambda$ & $\sigma_{\chi N}$ & $\Lambda$ & $\sigma_{\chi N}$ \\
(\GeVns{}) & (\GeVns{}) & (cm$^2$) & (\GeVns{}) & (cm$^2$) & (\GeVns{}) & (cm$^2$) & (\GeVns{}) & (cm$^2$) \\
\hline
1 & 572 & 2.4$\times10^{-39}$ & 626 & 1.7$\times10^{-39}$ & 522 & 3.5$\times10^{-39}$ & 516 & 3.7$\times10^{-39}$ \\
10 & 570 & 7.8$\times10^{-39}$ & 624 & 5.4$\times10^{-39}$ & 520 & 1.1$\times10^{-38}$ & 514 & 1.2$\times10^{-38}$ \\
100 & 571 & 9.1$\times10^{-39}$ & 625 & 6.3$\times10^{-39}$ & 521 & 1.3$\times10^{-38}$ & 510 & 1.4$\times10^{-38}$ \\
200 & 554 & 1.0$\times10^{-38}$ & 606 & 7.2$\times10^{-39}$ & 505 & 1.5$\times10^{-38}$ & 492 & 1.7$\times10^{-38}$ \\
300 & 533 & 1.2$\times10^{-38}$ & 583 & 8.5$\times10^{-39}$ & 486 & 1.7$\times10^{-38}$ & 471 & 2.0$\times10^{-38}$ \\
500 & 465 & 2.1$\times10^{-38}$ & 509 & 1.5$\times10^{-38}$ & 425 & 3.0$\times10^{-38}$ & 413 & 3.4$\times10^{-38}$ \\
1000 & 281 & 1.6$\times10^{-37}$ & 308 & 1.1$\times10^{-37}$ & 257 & 2.3$\times10^{-37}$ & 247 & 2.6$\times10^{-37}$ \\
\end{scotch}
\label{tab:DM_Limits_D5}
\end{table*}

\begin{table*}[h!tb]
\centering
\topcaption{Expected and observed 90\% \CL upper limits on the DM-nucleon cross section $\sigma_{\chi N}$ and
        effective cutoff scale $\Lambda$ for operator D8.}
\begin{scotch}{ccccccccc}
$m_\chi$ & \multicolumn{2}{c}{Expected} & \multicolumn{2}{c}{Expected$-1\sigma$}
       & \multicolumn{2}{c}{Expected$+1\sigma$} & \multicolumn{2}{c}{Observed} \\
\hline
      & $\Lambda$ & $\sigma_{\chi N}$ & $\Lambda$ & $\sigma_{\chi N}$ & $\Lambda$ & $\sigma_{\chi N}$ & $\Lambda$ & $\sigma_{\chi N}$ \\
(\GeVns{}) & (\GeVns{}) & (cm$^2$) & (\GeVns{}) & (cm$^2$) & (\GeVns{}) & (cm$^2$) & (\GeVns{}) & (cm$^2$) \\
\hline
1 & 574 & 8.8$\times10^{-41}$ & 627 & 6.1$\times10^{-41}$ & 523 & 1.3$\times10^{-40}$ & 518 & 1.3$\times10^{-40}$ \\
10 & 567 & 2.9$\times10^{-40}$ & 620 & 2.0$\times10^{-40}$ & 517 & 4.2$\times10^{-40}$ & 511 & 4.4$\times10^{-40}$ \\
100 & 555 & 3.7$\times10^{-40}$ & 607 & 2.6$\times10^{-40}$ & 507 & 5.3$\times10^{-40}$ & 498 & 5.7$\times10^{-40}$ \\
200 & 522 & 4.8$\times10^{-40}$ & 570 & 3.3$\times10^{-40}$ & 476 & 6.9$\times10^{-40}$ & 467 & 7.5$\times10^{-40}$ \\
300 & 479 & 6.7$\times10^{-40}$ & 524 & 4.7$\times10^{-40}$ & 437 & 9.7$\times10^{-40}$ & 425 & 1.1$\times10^{-39}$ \\
500 & 386 & 1.6$\times10^{-39}$ & 422 & 1.1$\times10^{-39}$ & 352 & 2.3$\times10^{-39}$ & 340 & 2.7$\times10^{-39}$ \\
1000 & 199 & 2.3$\times10^{-38}$ & 218 & 1.6$\times10^{-38}$ & 182 & 3.3$\times10^{-38}$ & 176 & 3.7$\times10^{-38}$ \\
\end{scotch}
\label{tab:DM_Limits_D8}
\end{table*}

\begin{table*}[htb!]
\centering
\topcaption{Expected and observed 90\% \CL upper limits on the DM-nucleon cross section $\sigma_{\chi N}$ and
        effective cutoff scale $\Lambda$ for operator D9.}
\begin{scotch}{ccccccccc}
$m_\chi$ & \multicolumn{2}{c}{Expected} & \multicolumn{2}{c}{Expected$-1\sigma$}
       & \multicolumn{2}{c}{Expected$+1\sigma$} & \multicolumn{2}{c}{Observed} \\
\hline
      & $\Lambda$ & $\sigma_{\chi N}$ & $\Lambda$ & $\sigma_{\chi N}$ & $\Lambda$ & $\sigma_{\chi N}$ & $\Lambda$ & $\sigma_{\chi N}$ \\
(\GeVns{}) & (\GeVns{}) & (cm$^2$) & (\GeVns{}) & (cm$^2$) & (\GeVns{}) & (cm$^2$) & (\GeVns{}) & (cm$^2$) \\
\hline
1 & 2139 & 4.5$\times10^{-43}$ & 2339 & 3.2$\times10^{-43}$ & 1951 & 6.5$\times10^{-43}$ & 1879 & 7.6$\times10^{-43}$ \\
10 & 2137 & 1.4$\times10^{-42}$ & 2337 & 1.0$\times10^{-42}$ & 1950 & 2.1$\times10^{-42}$ & 1864 & 2.5$\times10^{-42}$ \\
100 & 2102 & 1.8$\times10^{-42}$ & 2299 & 1.3$\times10^{-42}$ & 1918 & 2.6$\times10^{-42}$ & 1865 & 2.9$\times10^{-42}$ \\
200 & 2000 & 2.2$\times10^{-42}$ & 2187 & 1.5$\times10^{-42}$ & 1825 & 3.2$\times10^{-42}$ & 1772 & 3.6$\times10^{-42}$ \\
300 & 1863 & 2.9$\times10^{-42}$ & 2038 & 2.1$\times10^{-42}$ & 1700 & 4.2$\times10^{-42}$ & 1650 & 4.8$\times10^{-42}$ \\
500 & 1562 & 6.0$\times10^{-42}$ & 1708 & 4.2$\times10^{-42}$ & 1425 & 8.6$\times10^{-42}$ & 1395 & 9.4$\times10^{-42}$ \\
1000 & 886 & 5.8$\times10^{-41}$ & 969 & 4.0$\times10^{-41}$ & 809 & 8.3$\times10^{-41}$ & 790 & 9.1$\times10^{-41}$ \\
\end{scotch}
\label{tab:DM_Limits_D9}
\end{table*}

\begin{table*}[htb!]
\centering
\topcaption{Expected and observed 90\% \CL upper limits on the DM-nucleon cross section $\sigma_{\chi N}$ and
        effective cutoff scale $\Lambda$ for operator C3.}
\begin{scotch}{ccccccccc}
$m_\chi$ & \multicolumn{2}{c}{Expected} & \multicolumn{2}{c}{Expected$-1\sigma$}
       & \multicolumn{2}{c}{Expected$+1\sigma$} & \multicolumn{2}{c}{Observed} \\
\hline
      & $\Lambda$ & $\sigma_{\chi N}$ & $\Lambda$ & $\sigma_{\chi N}$ & $\Lambda$ & $\sigma_{\chi N}$ & $\Lambda$ & $\sigma_{\chi N}$ \\
(\GeVns{}) & (\GeVns{}) & (cm$^2$) & (\GeVns{}) & (cm$^2$) & (\GeVns{}) & (cm$^2$) & (\GeVns{}) & (cm$^2$) \\
\hline
1 & 403 & 4.0$\times10^{-38}$ & 440 & 2.8$\times10^{-38}$ & 367 & 5.7$\times10^{-38}$ & 363 & 6.1$\times10^{-38}$ \\
10 & 403 & 1.2$\times10^{-37}$ & 441 & 8.7$\times10^{-38}$ & 368 & 1.8$\times10^{-37}$ & 364 & 1.9$\times10^{-37}$ \\
100 & 396 & 1.6$\times10^{-37}$ & 433 & 1.1$\times10^{-37}$ & 361 & 2.3$\times10^{-37}$ & 356 & 2.4$\times10^{-37}$ \\
200 & 365 & 2.2$\times10^{-37}$ & 399 & 1.5$\times10^{-37}$ & 333 & 3.2$\times10^{-37}$ & 326 & 3.5$\times10^{-37}$ \\
300 & 335 & 3.1$\times10^{-37}$ & 366 & 2.2$\times10^{-37}$ & 305 & 4.5$\times10^{-37}$ & 297 & 5.0$\times10^{-37}$ \\
500 & 273 & 7.0$\times10^{-37}$ & 299 & 4.9$\times10^{-37}$ & 250 & 1.0$\times10^{-36}$ & 241 & 1.2$\times10^{-36}$ \\
1000 & 143 & 9.5$\times10^{-36}$ & 156 & 6.6$\times10^{-36}$ & 130 & 1.4$\times10^{-35}$ & 125 & 1.6$\times10^{-35}$ \\
\end{scotch}
\label{tab:DM_Limits_C3}
\end{table*}

\begin{figure}[htb!]
\centering
\includegraphics[width=0.48\textwidth]{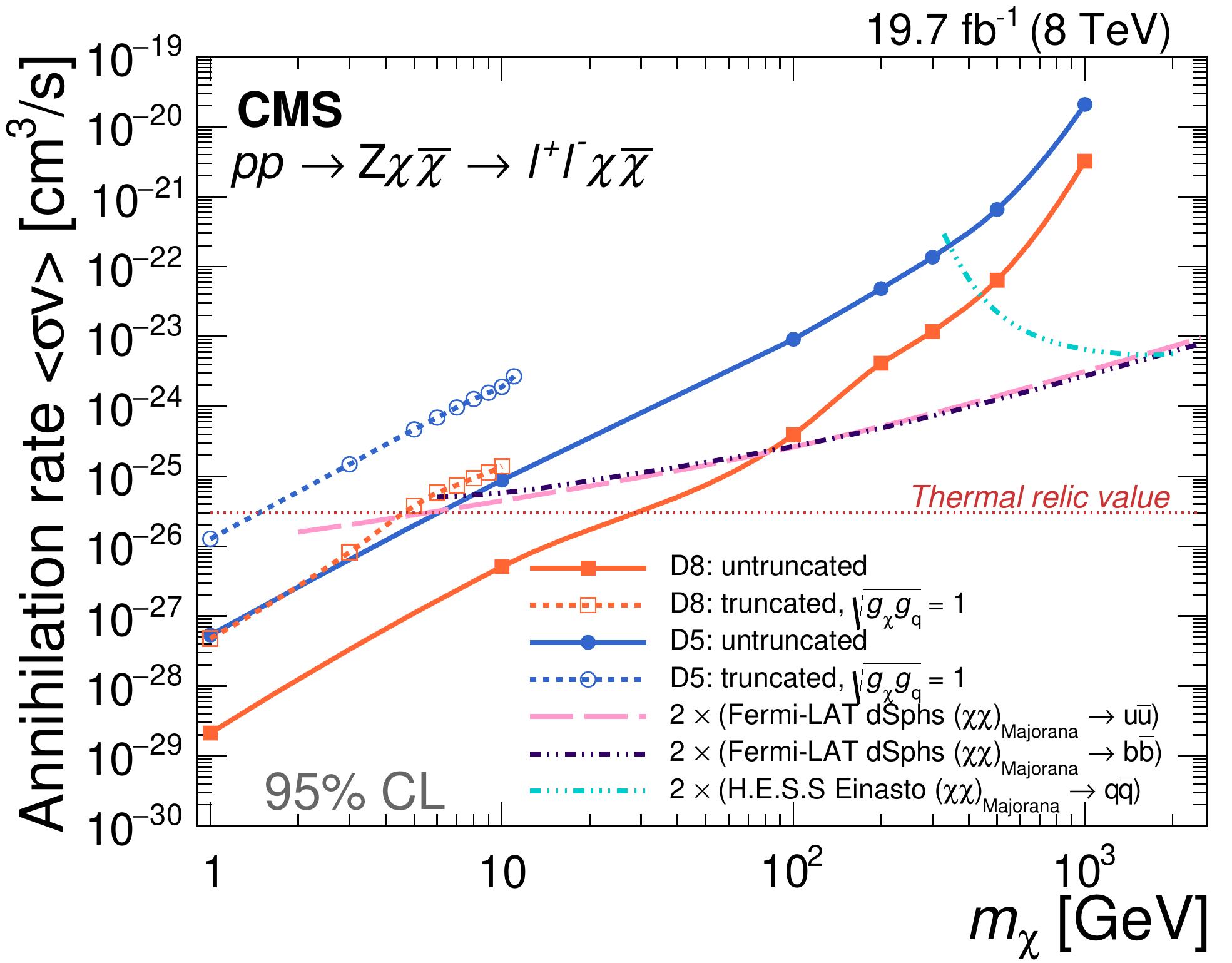}
\caption{The 95\%~\CL upper limits on the DM annihilation rate $\langle\sigma v\rangle$ for $\chi \overline{\chi} \to \PQq \PAQq$  as a function
	of the DM particle mass for vector (D5) and axial-vector (D8) couplings of Dirac fermion DM.
	A 100\% branching fraction of DM annihilating to quarks is assumed.
	Indirect search experimental results from H.E.S.S~\cite{HESS:2015cda} and
	Fermi-LAT~\cite{Ackermann:2013yva} are also plotted.
	The value required for DM particles to account for the relic abundance is labeled
	``Thermal relic value'' and is shown as a red dotted line.
	The truncated limits for D5 and D8 with
        $\sqrt{g_\cPq g_\chi}=1$ are presented with dashed lines in same shade as the untruncated ones.
	}
\label{fig:DM-Anni}
\end{figure}

Figure~\ref{fig:DM-Anni} shows the limits from operators D5 and D8 translated into upper
limits on the DM annihilation rate $\langle\sigma v\rangle$ relevant to indirect astrophysical searches~\cite{Fox:2011pm},
in which $\sigma$ is the annihilation cross section,
$v$ is the relative velocity of the annihilating particles,
and the quantity $\langle\sigma v\rangle$ is averaged over the distribution of the DM velocity.
In this paper, a particular astrophysical environment with $\langle v^2\rangle = 0.24$ is considered,
which corresponds to the epoch of the early universe when DM froze out, producing the thermal relic abundance.
A 100\% branching fraction of DM annihilating to quarks is assumed.
The corresponding truncated limits for D5 and D8 with coupling
$\sqrt{g_\cPq g_\chi}=1$ are also presented with dashed lines in same shade as the untruncated ones.
The value required for DM particles to make up the relic abundance is labeled ``Thermal relic value''
and is shown as a red dotted line.
With this constraint on the annihilation rate, we can conclude that
Dirac fermion DM is ruled out at 95\% \CL for $m_\chi < 6\GeV$ in the case of
vector coupling and $m_\chi < 30\GeV$ in the case of axial-vector coupling.
Indirect search results from H.E.S.S~\cite{HESS:2015cda} and
Fermi-LAT~\cite{Ackermann:2013yva} are also shown for comparison.
These results have been multiplied by a factor of 2 since they assume Majorana rather than Dirac fermions.

\subsection{Unparticle interpretation}

\begin{figure}[htb!]
\centering
\includegraphics[width=0.48\textwidth]{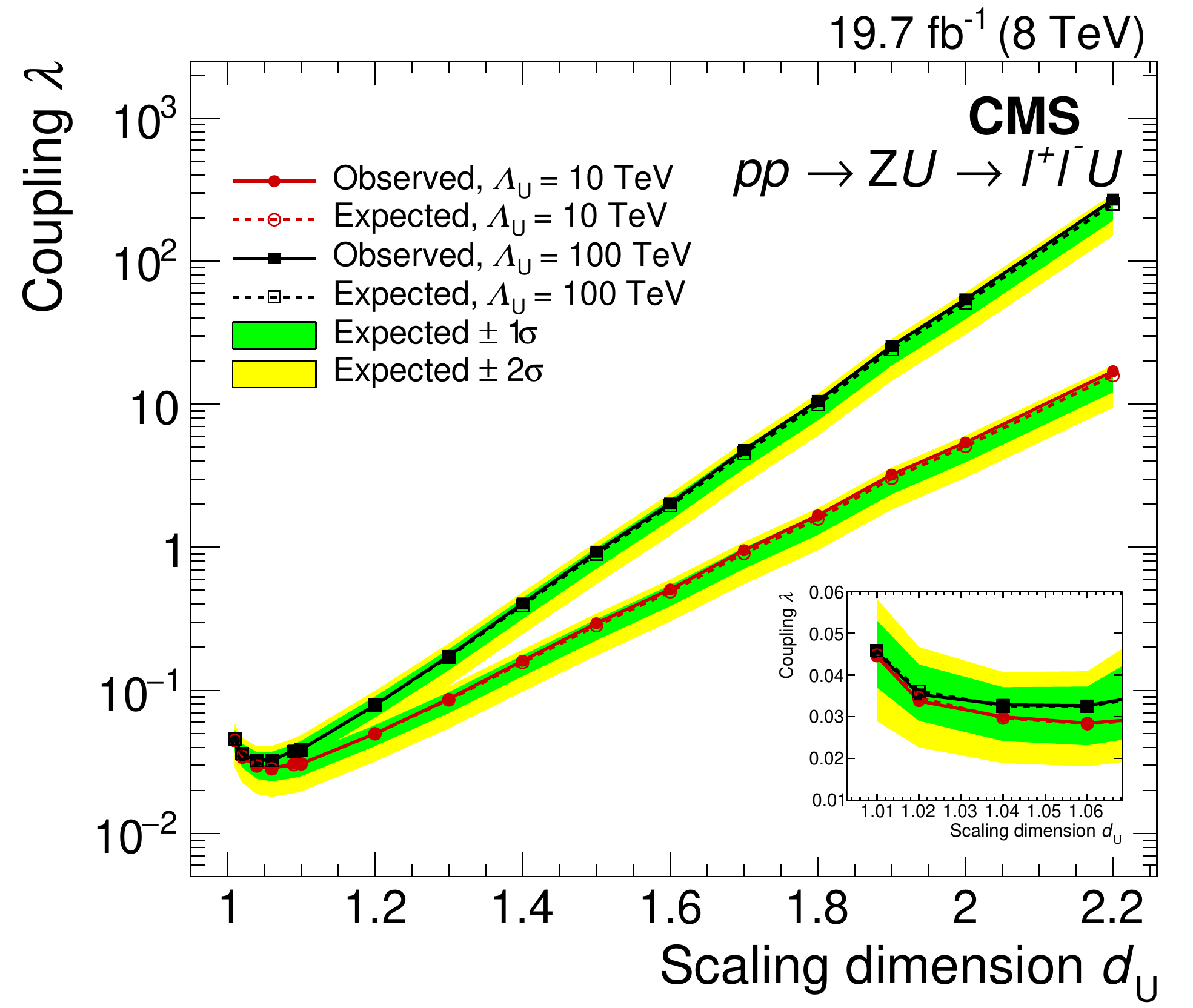}
\includegraphics[width=0.48\textwidth]{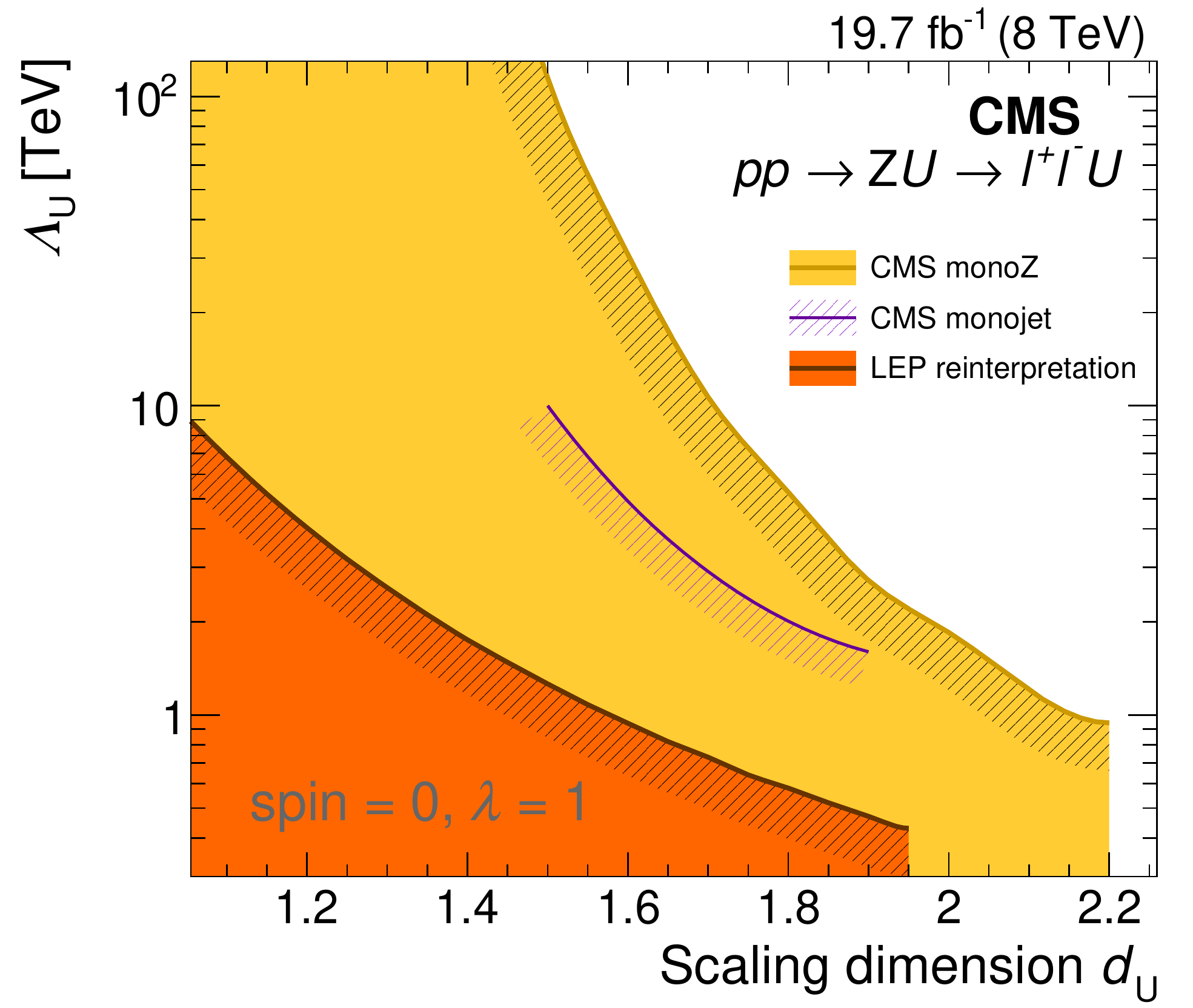}
\caption{\cmsLLeft: 95\% \CL upper limits on the coupling $\lambda$ between the unparticle and SM fields with fixed effective cutoff scales
	 $\Lambda_\mathcal{U}=10$ and 100\TeV. The plot inserted provides an expanded view of the limits at low scaling dimension.
	  \cmsRRight: 95\% \CL lower limits on unparticle effective cutoff scale $\Lambda_\mathcal{U}$ with a fixed coupling $\lambda=1$.
	  The results from CMS monojet~\cite{Khachatryan:2014rra} and reinterpretation of LEP searches~\cite{Kathrein:2010ej}
	  are also shown for comparison. The excluded region is indicated by the shading.
	}
\label{fig:unparticleLimits}
\end{figure}

In the scenario of the unparticle model, the 95\% \CL upper limits on the coupling constant $\lambda$ between
the unparticle and the SM fields with fixed effective cutoff scales $\Lambda_\mathcal{U}=10$\TeV and 100\TeV, as functions of the scaling
dimension $d_\mathcal{U}$, are shown on the left of Fig.~\ref{fig:unparticleLimits}. The right-hand plot of Fig.~\ref{fig:unparticleLimits}
presents 95\% \CL lower limits on the effective cutoff scale $\Lambda_\mathcal{U}$ with a fixed coupling $\lambda=1$
and compares the result with the limits obtained from
the CMS monojet search~\cite{Khachatryan:2014rra} and reinterpretation of LEP searches~\cite{Kathrein:2010ej}.
The search presented in this paper (labeled ``monoZ'') gives the most stringent limits.
Tables~\ref{tab:Limits_lambda_Unparticle10TeV} and~\ref{tab:Limits_lambda_Unparticle100TeV} show the
95\% \CL upper limits on the coupling $\lambda$
between unparticles and the SM fields for values of the scaling dimension $d_\mathcal{U}$ in the range
from 1.01 to 2.2, and fixed effective cutoff scales of 10\TeV and 100\TeV.
Lower limits at 95\% \CL on the effective cutoff scale $\Lambda_\mathcal{U}$ are given in Table~\ref{tab:Limits_LambdaU_Unparticle},
for $d_\mathcal{U}$ in the range from 1.6 to 2.2 and a fixed coupling $\lambda=1$.

\begin{table}[h!tb]
\centering
\topcaption{Expected and observed 95\% \CL upper limits on the coupling $\lambda$ between
	unparticles and the SM fields,
	for values of $d_\mathcal{U}$ in the range from 1.01 to 2.20 and a fixed effective cutoff
	scale $\Lambda_\mathcal{U}=10\TeV$.
	}
\begin{scotch}{.....}
\multicolumn{1}{l}{$d_\mathcal{U}$} & \multicolumn{4}{c}{$\lambda$} \\ \cline{2-5}
 & \multicolumn{1}{l}{Expected} & \multicolumn{1}{l}{Expected$-1\sigma$} & \multicolumn{1}{l}{Expected$+1\sigma$} & \multicolumn{1}{l}{Observed} \\
\hline
1.01 & 0.045 & 0.038 & 0.053 & 0.044 \\
1.02 & 0.035 & 0.030 & 0.042 & 0.034 \\
1.04 & 0.029 & 0.025 & 0.035 & 0.030 \\
1.06 & 0.028 & 0.024 & 0.033 & 0.028 \\
1.09 & 0.030 & 0.025 & 0.035 & 0.031 \\
1.10 & 0.031 & 0.026 & 0.036 & 0.031 \\
1.30 & 0.085 & 0.072 & 0.100 & 0.087 \\
1.50 & 0.273 & 0.232 & 0.322 & 0.295 \\
1.70 & 0.864 & 0.734 & 1.018 & 0.956 \\
1.90 & 2.86 & 2.43 & 3.37 & 3.22 \\
2.20 & 14.8 & 12.6 & 17.4 & 17.0 \\
\end{scotch}
\label{tab:Limits_lambda_Unparticle10TeV}
\end{table}

\begin{table}[h!tb]
\centering
\topcaption{Expected and observed 95\% \CL upper limits on the coupling $\lambda$ between
	unparticles and the SM fields,
	for values of $d_\mathcal{U}$ in the range from 1.01 to 2.20 and a fixed effective cutoff
	scale $\Lambda_\mathcal{U}=100\TeV$.
	}
\begin{scotch}{.....}
\multicolumn{1}{l}{$d_\mathcal{U}$} & \multicolumn{4}{c}{$\lambda$} \\ \cline{2-5}
 & \multicolumn{1}{l}{Expected} & \multicolumn{1}{l}{Expected$-1\sigma$} & \multicolumn{1}{l}{Expected$+1\sigma$} & \multicolumn{1}{l}{Observed} \\
\hline
1.01 & 0.046 & 0.039 & 0.054 & 0.045 \\
1.02 & 0.037 & 0.031 & 0.044 & 0.035 \\
1.04 & 0.032 & 0.027 & 0.038 & 0.033 \\
1.06 & 0.032 & 0.027 & 0.038 & 0.033 \\
1.09 & 0.037 & 0.031 & 0.043 & 0.038 \\
1.10 & 0.039 & 0.033 & 0.046 & 0.039 \\
1.30 & 0.169 & 0.143 & 0.199 & 0.174 \\
1.50 & 0.864 & 0.734 & 1.018 & 0.933 \\
1.60 & 1.88 & 1.60 & 2.22 & 2.02 \\
1.70 & 4.33 & 3.68 & 5.10 & 4.79 \\
1.90 & 22.7 & 19.3 & 26.8 & 25.6 \\
2.20 & 235 & 199 & 276 & 270 \\
\end{scotch}
\label{tab:Limits_lambda_Unparticle100TeV}
\end{table}

\begin{table}[h!tb]
\centering
\topcaption{Expected and observed 95\% \CL lower limits on the effective cutoff scale
	$\Lambda_\mathcal{U}$ for values of $d_\mathcal{U}$ in the range from 1.60 to 2.20 and
	a fixed coupling $\lambda=1$.
        }
\begin{scotch}{c....}
\multicolumn{1}{l}{$d_\mathcal{U}$} & \multicolumn{4}{c}{$\Lambda_\mathcal{U}$ (TeV)} \\ \cline{2-5}
 & \multicolumn{1}{l}{Expected} & \multicolumn{1}{l}{Expected$-1\sigma$} & \multicolumn{1}{l}{Expected$+1\sigma$} & \multicolumn{1}{l}{Observed} \\
\hline
1.50 & 134 & 186 & 96.4 & 115 \\
1.60 & 34.8 & 45.7 & 26.5 & 30.9 \\
1.70 & 12.3 & 15.6 & 9.75 & 10.7 \\
1.80 & 6.08 & 7.45 & 4.95 & 5.25 \\
1.90 & 3.11 & 3.72 & 2.59 & 2.72 \\
2.00 & 2.09 & 2.46 & 1.77 & 1.85 \\
2.20 & 1.06 & 1.21 & 0.92 & 0.94 \\
\end{scotch}
\label{tab:Limits_LambdaU_Unparticle}
\end{table}

\subsection{Model-independent limits}

\begin{figure}[htb!]
\centering
\includegraphics[width=0.48\textwidth]{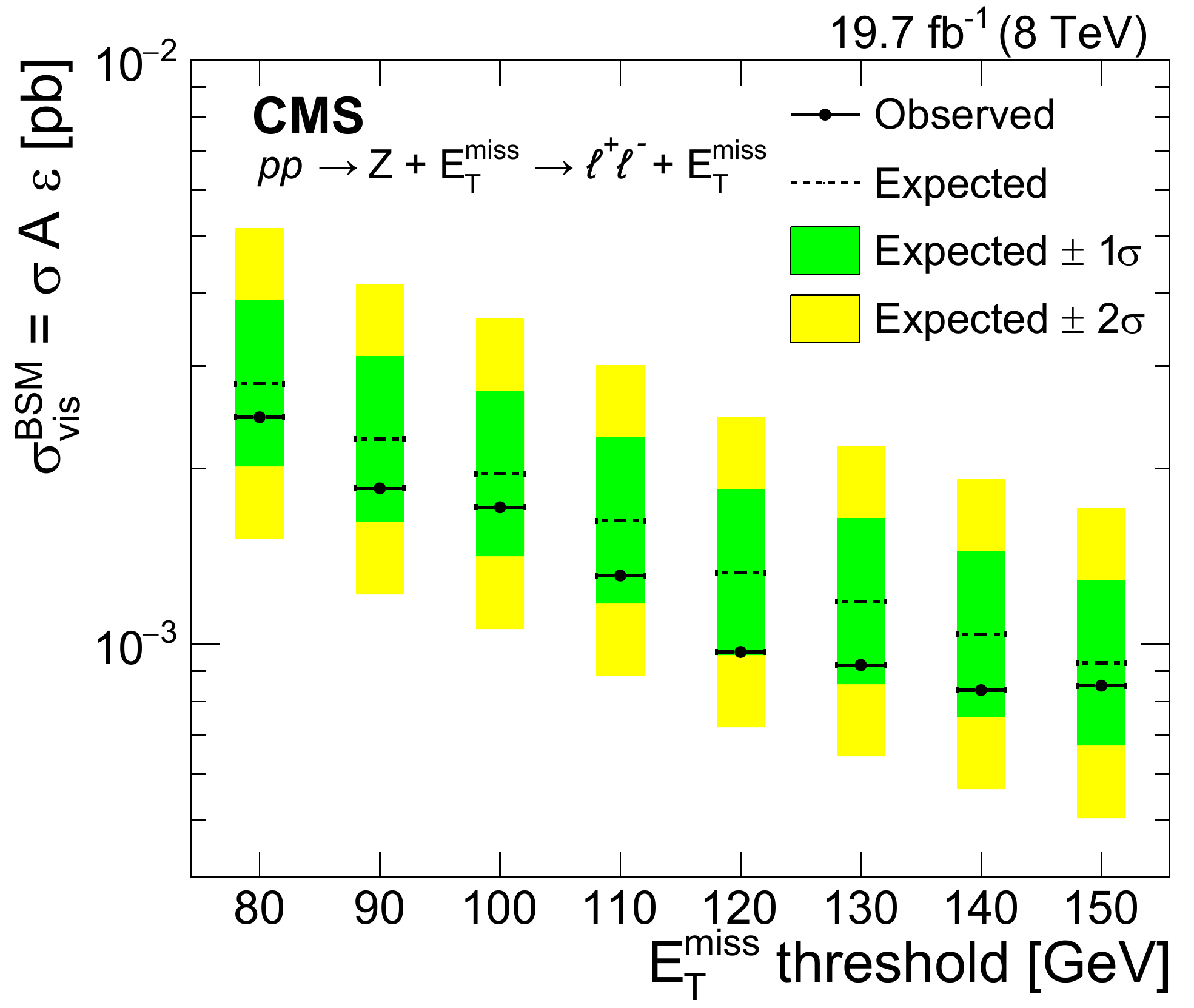}
\caption{The model-independent upper limits at 95\% \CL on the visible cross section ($\sigma\, A\,\epsilon$) for BSM production of events, as a function of \ETm threshold.}
\label{fig:modelIndepLimits}
\end{figure}

As an alternative to the interpretation of the results in specific models, a single-bin analysis is applied to obtain model-independent
expected and observed 95\% \CL upper limits on the visible cross section $\sigma_\text{vis}^\mathrm{BSM}$
for beyond the standard model (BSM) physics processes. The limits as a function of \ETm thresholds are shown in Fig.~\ref{fig:modelIndepLimits}.
With a \ETm threshold of 80\,(150)\GeV, we exclude the visible cross section $\sigma_\text{vis}^\mathrm{BSM}>2.5\,(0.85)$\unit{fb}.
Table~\ref{tab:model_indep} shows the total SM background predictions for the numbers of events passing the selection requirements,
for different \ETm thresholds, compared with the observed numbers of events.
The 95\% \CL expected and observed upper limits for the contribution of events from BSM sources are also shown.

\begin{table*}[!htb]
\centering
\topcaption{Total SM background predictions for the numbers of events passing the selection requirements,
	for different \ETm thresholds, compared with the observed numbers of events.
	The listed uncertainties include both statistical and systematic components.
	The 95\% \CL observed and expected upper limits for the contribution of events from BSM sources are also shown.
	The ${\pm}1\sigma$ and ${\pm}2\sigma$ excursions from expected limits are also given.}
\begin{scotch}{lcccccccc}
\ETm (\GeVns{}) threshold & 80 & 90 & 100 & 110 & 120 & 130 & 140 & 150 \\
\hline

Total SM    &  263 &  193 &  150 &  117 &  90.5 &  72.5 &  59.2 &  45.1 \\
Total uncertainty &   $\pm$30  &   $\pm$24  &   $\pm$20  &   $\pm$16  &   $\pm$13  &   $\pm$12  &   $\pm$9.6  &   $\pm$7.6 \\
Data	    &    244 &    172 &    141 &    104 &    74 &    61 &    50 &    43\\
\hline
Obs. upper limit &  48.3 &  36.5 &  33.8 &  25.9 &  19.1 &  18.2 &  16.5 &  16.7\\
Exp. upper limit +2$\sigma$ &  102 &  81.7 &  71.3 &  59.2 &  48.3 &  43.1 &  37.9 &  33.8\\
Exp. upper limit +1$\sigma$ &  76.5 &  61.5 &  53.7 &  44.6 &  36.4 &  32.4 &  28.5 &  25.4\\
Exp. upper limit &  55.1 &  44.3 &  38.6 &  32.1 &  26.2 &  23.4 &  20.5 &  18.3\\
Exp. upper limit -1$\sigma$ &  39.7 &  32.0 &  27.9 &  23.2 &  18.9 &  16.9 &  14.8 &  13.2\\
Exp. upper limit -2$\sigma$ &  29.9 &  24.0 &  21.0 &  17.4 &  14.2 &  12.7 &  11.1 &   9.90\\
\end{scotch}
\label{tab:model_indep}
\end{table*}

\section{Summary}
\label{sec:summary}
A search for evidence for particle dark matter and unparticle production at the LHC has been performed in events
containing two charged leptons, consistent with the decay of a \Z boson, and large missing transverse momentum.
The study is based on a data set corresponding to an integrated luminosity of 19.7\fbinv
of pp collisions collected by the CMS detector at a center-of-mass energy of 8\TeV.
The results are consistent with the expected standard model contributions.
These results are interpreted in two scenarios for physics beyond the standard model: dark matter and unparticles.
Model-independent 95\% confidence level upper limits are also set on contributions to the visible $\Z$+\ETm cross section from sources beyond the standard model.
Upper limits at 90\% confidence level are set on the DM-nucleon scattering
cross sections as a function of DM particle mass for both spin-dependent and spin-independent cases.
Limits are also set on the DM annihilation rate assuming a branching fraction of 100\% for annihilation to quarks,
and on the effective cutoff scale.
In addition, the most stringent limits to date at 95\% confidence level on the coupling between unparticles and the standard model fields
as well as the effective cutoff scale as a function of the unparticle scaling dimension
are obtained in this analysis.

\begin{acknowledgments}
\hyphenation{Bundes-ministerium Forschungs-gemeinschaft Forschungs-zentren} We congratulate our colleagues in the CERN accelerator departments for the excellent performance of the LHC and thank the technical and administrative staffs at CERN and at other CMS institutes for their contributions to the success of the CMS effort. In addition, we gratefully acknowledge the computing centers and personnel of the Worldwide LHC Computing Grid for delivering so effectively the computing infrastructure essential to our analyses. Finally, we acknowledge the enduring support for the construction and operation of the LHC and the CMS detector provided by the following funding agencies: the Austrian Federal Ministry of Science, Research and Economy and the Austrian Science Fund; the Belgian Fonds de la Recherche Scientifique and Fonds voor Wetenschappelijk Onderzoek; the Brazilian Funding Agencies (CNPq, CAPES, FAPERJ, and FAPESP); the Bulgarian Ministry of Education and Science; CERN; the Chinese Academy of Sciences, Ministry of Science and Technology, and National Natural Science Foundation of China; the Colombian Funding Agency (COLCIENCIAS); the Croatian Ministry of Science, Education and Sport, and the Croatian Science Foundation; the Research Promotion Foundation, Cyprus; the Ministry of Education and Research, Estonian Research Council via IUT23-4 and IUT23-6 and European Regional Development Fund, Estonia; the Academy of Finland, Finnish Ministry of Education and Culture, and Helsinki Institute of Physics; the Institut National de Physique Nucl\'eaire et de Physique des Particules/CNRS and Commissariat \`a l'\'Energie Atomique et aux \'Energies Alternatives/CEA, France; the Bundesministerium f\"ur Bildung und Forschung, Deutsche Forschungsgemeinschaft, and Helmholtz-Gemeinschaft Deutscher Forschungszentren, Germany; the General Secretariat for Research and Technology, Greece; the National Scientific Research Foundation, and National Innovation Office, Hungary; the Department of Atomic Energy and the Department of Science and Technology, India; the Institute for Studies in Theoretical Physics and Mathematics, Iran; the Science Foundation, Ireland; the Istituto Nazionale di Fisica Nucleare, Italy; the Ministry of Science, ICT and Future Planning, and National Research Foundation (NRF), Republic of Korea; the Lithuanian Academy of Sciences; the Ministry of Education, and University of Malaya, Malaysia; the Mexican Funding Agencies (CINVESTAV, CONACYT, SEP, and UASLP-FAI); the Ministry of Business, Innovation and Employment, New Zealand; the Pakistan Atomic Energy Commission; the Ministry of Science and Higher Education and the National Science Centre, Poland; the Funda\c{c}\~ao para a Ci\^encia e a Tecnologia, Portugal; JINR, Dubna, the Ministry of Education and Science of the Russian Federation, the Federal Agency of Atomic Energy of the Russian Federation, Russian Academy of Sciences, and the Russian Foundation for Basic Research; the Ministry of Education, Science and Technological Development of Serbia; the Secretar\'{\i}a de Estado de Investigaci\'on, Desarrollo e Innovaci\'on and Programa Consolider-Ingenio 2010, Spain; the Swiss Funding Agencies (ETH Board, ETH Zurich, PSI, SNF, UniZH, Canton Zurich, and SER); the Ministry of Science and Technology, Taipei; the Thailand Center of Excellence in Physics, the Institute for the Promotion of Teaching Science and Technology of Thailand, Special Task Force for Activating Research, and the National Science and Technology Development Agency of Thailand; the Scientific and Technical Research Council of Turkey and Turkish Atomic Energy Authority; the National Academy of Sciences of Ukraine and State Fund for Fundamental Researches, Ukraine; the Science and Technology Facilities Council, United Kingdom; the US Department of Energy and the US National Science Foundation.

Individuals have received support from the Marie-Curie program and the European Research Council and EPLANET (European Union); the Leventis Foundation; the A. P. Sloan Foundation; the Alexander von Humboldt Foundation; the Belgian Federal Science Policy Office; the Fonds pour la Formation \`a la Recherche dans l'Industrie et dans l'Agriculture (FRIA-Belgium); the Agentschap voor Innovatie door Wetenschap en Technologie (IWT-Belgium); the Ministry of Education, Youth and Sports (MEYS) of the Czech Republic; the Council of Science and Industrial Research, India; the HOMING PLUS program of the Foundation for Polish Science, cofinanced by the European Union, Regional Development Fund; the OPUS program of the National Science Center (Poland); the Compagnia di San Paolo (Torino); MIUR Project No.~20108T4XTM (Italy); the Thalis and Aristeia programs cofinanced by EU-ESF and the Greek NSRF; the National Priorities Research Program by Qatar National Research Fund; the Rachadapisek Sompot Fund for Postdoctoral Fellowship, Chulalongkorn University (Thailand); and the Welch Foundation, Contract No.~C-1845.
\end{acknowledgments}

\bibliography{auto_generated}

\cleardoublepage \appendix\section{The CMS Collaboration \label{app:collab}}\begin{sloppypar}\hyphenpenalty=5000\widowpenalty=500\clubpenalty=5000\textbf{Yerevan Physics Institute,  Yerevan,  Armenia}\\*[0pt]
V.~Khachatryan, A.M.~Sirunyan, A.~Tumasyan
\vskip\cmsinstskip
\textbf{Institut f\"{u}r Hochenergiephysik der OeAW,  Wien,  Austria}\\*[0pt]
W.~Adam, E.~Asilar, T.~Bergauer, J.~Brandstetter, E.~Brondolin, M.~Dragicevic, J.~Er\"{o}, M.~Flechl, M.~Friedl, R.~Fr\"{u}hwirth\cmsAuthorMark{1}, V.M.~Ghete, C.~Hartl, N.~H\"{o}rmann, J.~Hrubec, M.~Jeitler\cmsAuthorMark{1}, V.~Kn\"{u}nz, A.~K\"{o}nig, M.~Krammer\cmsAuthorMark{1}, I.~Kr\"{a}tschmer, D.~Liko, T.~Matsushita, I.~Mikulec, D.~Rabady\cmsAuthorMark{2}, B.~Rahbaran, H.~Rohringer, J.~Schieck\cmsAuthorMark{1}, R.~Sch\"{o}fbeck, J.~Strauss, W.~Treberer-Treberspurg, W.~Waltenberger, C.-E.~Wulz\cmsAuthorMark{1}
% ...
\vskip\cmsinstskip
% ...
1:~~Also at Vienna University of Technology, Vienna, Austria\\
2:~~Also at CERN, European Organization for Nuclear Research, Geneva, Switzerland\\
% ...

\end{sloppypar}
\end{document}